\documentclass[preprint]{elsarticle}
\usepackage{lineno,hyperref}
\modulolinenumbers[5]
\journal{}
\usepackage{setspace}

\renewcommand{\baselinestretch}{1.1} 
\usepackage{geometry}

\newgeometry{vmargin={25mm}, hmargin={20mm,20mm}}
\usepackage{color}
\usepackage{amsmath}

\usepackage{amsmath}
\usepackage{lineno,hyperref}
\usepackage{subcaption}
\usepackage{graphicx}
\usepackage{amssymb,amsthm}
\usepackage{color}
\usepackage{xcolor}
\usepackage{tikz}
\usepackage{nicefrac}
\usepackage{xfrac}
\usepackage{tikz}
\usepackage{mathtools}

\modulolinenumbers[205]
\journal{Journal of \LaTeX\ Templates}
\usepackage{comment}
\usepackage{setspace}
\usepackage{hyperref}
\renewcommand{\baselinestretch}{1.1} 
\usepackage{geometry}
\usepackage[normalem]{ulem}
\newgeometry{vmargin={21mm}, hmargin={20mm,20mm}}

\begin{document}

\begin{frontmatter}

\title{Network inference combining mutual information rate and statistical tests}

\author[CA_mainaddress]{Chris G. Antonopoulos\corref{CAcorrespondingauthor}}
\cortext[CAcorrespondingauthor]{Corresponding author}
\ead{canton@essex.ac.uk}

\address[CA_mainaddress]{Department of Mathematical Sciences, University of Essex, Wivenhoe Park, UK}

\begin{abstract}
In this paper, we present a method that combines information-theoretical and statistical approaches to infer connectivity in complex networks using time-series data. The method is based on estimations of the Mutual Information Rate for pairs of time-series and on statistical significance tests for connectivity acceptance using the false discovery rate method for multiple hypothesis testing. We provide the mathematical background on Mutual Information Rate, discuss the statistical significance tests and the false discovery rate. Further on, we present results for correlated normal-variates data, coupled circle and coupled logistic maps, coupled Lorenz systems and coupled stochastic Kuramoto phase oscillators. Following up, we study the effect of noise on the presented methodology in networks of coupled stochastic Kuramoto phase oscillators and of coupling heterogeneity degree on networks of coupled circle maps. We show that the method can infer the correct number and pairs of connected nodes, by means of receiver operating characteristic curves. In the more realistic case of stochastic data, we demonstrate its ability to infer the structure of the initial connectivity matrices. The method is also shown to recover the initial connectivity matrices for dynamics on the nodes of Erd\H{o}s-R\'enyi and small-world networks with varying coupling heterogeneity in their connections. The highlight of the proposed methodology is its ability to infer the underlying network connectivity based solely on the recorded datasets.
\end{abstract}


\end{frontmatter}

%\linenumbers

%%%%%%%%%%%%%%%%%%%%%%%%%%%%%%%%%%%%%%%%%%%%%%
\section{Introduction}

Complex systems are ubiquitous in nature and are the subject of intense research over the last decades \cite{Thurneretal2018,Sayama2018}, which has led to the emergence of the new paradigm of complexity or ``complex systems'' science \cite{Sayama2018}. A complex system consists of many units which may interact in non-trivial ways, and whose aggregate behaviour is undetermined from the behaviour of the individual units \cite{Barrat2008,Sayama2018}. Prominent examples are organisms, the human brain, power grids, transportation and communication systems, ecological systems, etc., \cite{Sayama2018}. What is common in these systems is that their behaviour is intrinsically difficult to model due to dependencies, competitions, relationships, or other types of interactions among their units or between the system and its environment. What is fascinating is that they exhibit distinct properties that arise from these relationships, such as nonlinearity, emergence, self-organisation, spontaneous order, adaptation, feedback loops, etc. Due to their appearance in a variety of fields and disciplines, common approaches to study them have become prominent and topics of their own importance \cite{Sayama2018}. Interestingly, one can represent complex systems as a mathematical model of a graph that consists of nodes that represent the units in the system and edges that represent interactions among the nodes, a structure that is commonly referred to as a network.

Considering these units as nodes in the network, and their underlying physical interactions as links, a step forward in understanding complex systems is to infer the topological structure of their network representation, i.e., their connectivity. This is driven by the fact that in most cases, the connectivity is not known, and that one might only have access to data recorded on the nodes of the network, for example in experiments. Usually, the connectivity of the units is not known or is difficult to detect by physical methods due to large system-sizes. Consequently, it becomes clear that if one wants to study complex systems when only recorded data are available, one first has to infer the underlying structure, namely to establish whether there are undirected links among the pairs of nodes in the network.

These hurdles highlight the importance of the development of mathematical approaches to infer network connectivity using recorded data, i.e., methodologies that can reveal interconnectivity among nodes in a network. Depending on the research field, some of these approaches might be more applicable than others. For example, in \cite{HERMES_2013}, the authors developed HERMES, a Matlab software package with a collection of methods to infer functional and effective brain connectivity from neurophysiological data such as from multivariate electroencephalography (EEG) and magnetoencephalography (MEG) recordings. In \cite{Omony2014}, a review of methods and assessment of tools and techniques is discussed for biological network inference and in \cite{Sakkalis2011}, the author provides a review of advanced techniques for the estimation of brain connectivity measured with EEG and MEG. The authors in \cite{Olejarczyketal2017} compared connectivity analyses for resting state EEG data pointing to the advantages of nonlinear methods, and indicating a relationship between the flow of information and the level of synchronisation in the brain. In \cite{Jahanpour2014ImprovingAO}, it is discussed how to improve accuracy of complex network modelling using maximum likelihood estimation and expectation-maximisation.

Network inference in nonlinear systems, physics, engineering, neuroscience, biology, etc. has been extensively studied in recent years using, for example, cross-correlation (CC) or mutual information (MI) \cite{rubido2014exact,Tirabassi2015,butte}, recurrences \cite{mamen,mamen2010,Hempel2011}, functional, response and complex dynamics \cite{ta,Timme2007,Timme2011,casadiego}, Granger and multivariate Granger causality \cite{granger,Feng2010,Bressler2010,Barnettetal2014}, rank-based connectivity measures \cite{Leguia_et_al_2019}, rank statistics \cite{Chicharroetal2009}, mutual prediction \cite{Schiffetal1996}, phase transfer entropy \cite{Lobieretal2014}, etc. In \cite{PhysRevE.95.042302}, the authors present a method that can reconstruct directed and weighted topologies of networks of heterogeneous nonlinear oscillators. In \cite{Kralemann_2014}, a novel approach for the recovery of the directional connectivity of a small oscillator network by means of the phase dynamics reconstruction from multivariate time-series data is presented, where the authors use a triplet analysis instead of the traditional pairwise. In \cite{Yangetal2021}, the authors present a general framework for dynamic complex networks applicable for characterising dynamic complex networks. They discuss that whether a dynamic complex network evolves toward synchrony or deterioration and collapse can be determined by tracking ensemble entropy in time. They found that, in addition to misrepresenting the true network dynamics, static network structures do not differentiate in resolving network properties such as average path-lengths and degree distributions. Even though the progress made, network inference still presents open challenges. For one, any similarity measure usually results in a non-zero value \cite{steuer2002mutual,mamen2010} as, in finite datasets, persistent trends or deterministic recurrent oscillations result in spurious correlations \cite{Palus2007,Palusetal2011}. Also, nonlinearities and asymmetries in the structure or noise level of the dynamics can affect these methods \cite{schreiber2000measuring,Quirogaetal2000,Smirnovetal2003,Nalatoreetal2007}.

In this paper, we are using an information-theoretical methodology and statistical significance tests complemented by the false discovery rate (FDR) method for multiple hypothesis testing to infer connectivity in complex networks using only time-series data \cite{bianco2016successful,Gohetal2018}. The data are recorded on the nodes of the network. The equations of motions and initial connectivity matrices are only used to compute and record the solutions, and to compare the inferred with the initial connectivity matrices by means of receiver operating characteristic (ROC) curves for different parameter values. This allows to quantify how successful the network inference is. Our methodology is based on: (a) estimations of the Mutual Information Rate (MIR) for pairs of time-series (that represent pairs of nodes in the network) \cite{bianco2016successful,Gohetal2018} and (b) on statistical significance tests to accept or reject connectivity using the false discovery rate method for multiple hypothesis testing.

MIR is the rate of information exchanged per unit of time between pairs of time-series \cite{Baptistaetal2012,bianco2016successful,Gohetal2018} and is an appropriate measure to quantify the exchange of information in systems with correlation \cite{palus,schreiber2000measuring, Baptistaetal2012}. A normalised version of MIR has been shown to perform well in inferring the structure of networks in different cases of dynamics \cite{bianco2016successful} and, in financial and stock markets data \cite{Gohetal2018}. The authors in \cite{bianco2016successful} discuss how to calculate MIR for Markov partitions. Even though it is a very interesting approach, these partitions are in general difficult or impossible to find. The methodology presented here is applicable in the case where only recorded data are available. In \cite{rubido2014exact}, two inference methods based on CC and MI are compared and it was shown that, for the class of studied systems, when an abrupt change in the ordered CC or MI values exists, it is possible to infer, without errors, the underlying network topology from the time-series data, even in the presence of observational noise, non-identical units, and coupling heterogeneity. However, this approach might not be able to infer the initial topology in situations where the gaps in the ordered CC or MI values are absent or when there are more than one gaps as it happens for the financial and stock market data in \cite{Gohetal2018}. In these cases, a (unique) threshold cannot be chosen to infer the network structure \cite{rubido2014exact}. The approach presented here does not make use of thresholds: an inferred network will be computed based on the recorded data using pairwise computed MIR values and surrogate data to test the null hypotheses for connectivity using the false discovery rate method for multiple hypothesis testing.

In \cite{bianco2016successful}, the authors used the bin method and computed the probabilities for the estimation of MIR in partitions of e.g., equally sized cells on the probability space generated by a pair of variables or units $X$ and $Y$. Here, we use the same approach. This method leads to the overestimation of MIR for random systems or non-Markovian partitions \cite{steuer2002mutual,Herzeletal1994}. This is also the case here and is due to the finite resolution of non-Markovian partitions and to the finite length of the recorded time-series. According to \cite{steuer2002mutual,Herzeletal1994}, these errors are systematic and always present in the computation of MIR for arbitrary non-Markovian partitions. In \cite{bianco2016successful}, these systematic errors were mitigated by double-normalising MIR, first over the MIR values of the pairs of variables $X$ and $Y$ for a particular grid size, and then, over the number of grid sizes on the probability space. Here, we show that this double-normalisation is not necessary and that even though, the resulting MIR values are overestimated, they are enough to allow for network inference when combined with the use of surrogate data to test the null hypotheses for connectivity and the false discovery rate method for multiple hypothesis testing.

It is worth it to note that in this work, we are interested in identifying undirected connections among nodes as MIR is a symmetric quantity with respect to $X$ and $Y$, i.e., $\mbox{MIR}_{XY}=\mbox{MIR}_{YX}$. Thus, we assume throughout the paper only undirected connections, and that such connections are due to linear or nonlinear, dynamical interactions. This is tested by using surrogate data to test the null hypotheses for connectivity and the false discovery rate method for multiple hypothesis testing. Our results show that this combination of methods can be effective in inferring network connectivity.

The paper is organised as follows: In Sec. \ref{sec_theor_backgd}, we provide the mathematical background on information and network connectivity, on MI and MIR, on the estimation of the correlation-decay time and on network inference based on MIR and statistical significance tests using the false discovery rate for multiple hypothesis testing. In Sec. \ref{sec_results}, we report on the results for different types of processes and, for discrete and continuous dynamical systems, such as correlated normal-variates, coupled circle maps, coupled logistic maps, coupled Lorenz systems and coupled stochastic Kuramoto phase oscillators. We also study the effect of noise on the proposed methodology in networks of coupled stochastic Kuramoto phase oscillators and of the effect of coupling heterogeneity degree on the method in networks of coupled circle maps. In Sec. \ref{sec_disc}, we discuss our work in light of its advantages and disadvantages with respect to other approaches in the literature. Finally, in Sec. \ref{sec_concl}, we conclude our work and discuss possible extensions that would be worth studying further.

%%%%%%%%%%%%%%%%%%%%%%%%%%%%%%%%%%%%%%%%%%%%%%
\section{Theoretical background}\label{sec_theor_backgd}

\subsection{Information and network connectivity}

Complex systems are networks made of a number of units that interact with each other, typically in nonlinear ways \cite{Sayama2018}, without excluding the possibility of highly ordered processes that follow simple rules.\cite{Crutchfield2013}. These systems may arise and evolve through self-organisation, such that they are neither completely regular nor completely random, permitting the development of emergent behaviour at macroscopic scales \cite{Sayama2018}. In this context, a system, composed of units, can produce information that can be transferred among them through its connectivity \cite{paluvs1996coarse, schreiber2000measuring, Baptistaetal2012, marchiori2012energy, mandal2013maxwell, antonopoulos2014production, antonopoulos2015brain}, involving at least two interacting units. In general, these interactions may be direct or indirect and linear or nonlinear \cite{Crutchfield2013,Sayama2018}.

Particularly, the units of the system are coupled and are represented as nodes in a network. Each of these units, labelled by $i=1,\ldots,M$, can ``observe'' the behaviour of some neighbours $j=1,\ldots,J$ by a measurement function $h(X_j)$, where $X_j$ is the state of unit $j$. For example, each unit $i$ could be a R\"ossler oscillator, and the measurement function could be $h(X_j) = y_j$, meaning that the R\"ossler oscillators are coupled by their $y_j$ units. Here, we assume that the units are represented by recorded time-series, i.e., by the evolution of $y_j$ in time.

In what follows, we use MIR to estimate the amount of information exchanged per unit of time between pairs of units in a system, namely to determine whether the units are connected bidirectionally. In this context, bidirectional connectivity means an undirected connection among the units attributed to their interactions, either  linear or nonlinear. Particularly, MIR measures the amount of information exchanged per unit of time between two units. Its application to time-series data is of primordial importance and will be used to determine if an undirected connection  between pairs of units exists, and thus pairs of nodes in the network. In this framework, the strengths of the connections, given by their MIR values, will be estimated and compared with the MIR values of all other pairs of units in the network by means of statistical significance tests and the false discovery rate method for multiple hypothesis testing. The idea is to find out whether the connections among units, and thus among nodes exist due to their dynamics.

\subsection{Mutual information}

MI and MIR were originally introduced by Shannon in his seminal paper in 1948 \cite{Shannon1948}. In particular, the MI of two random variables, $X$ and $Y$, is a measure of their mutual dependence and quantifies the ``amount of information'' obtained about $X$ after observing $Y$ or vice versa. Formally, it is defined by \cite{Shannon1948,kullback1997}
\begin{equation}\label{IXY1}
I_{XY}(N)=H_X+H_Y-H_{XY},
\end{equation}
where $N$ is the number of random events in $X$ and $Y$. $H_X$ and $H_Y$ are the marginal entropies of $X$ and $Y$, respectively, i.e., their Shannon entropies, with $H_X$ given by
\begin{equation*}
H_X=-\sum_{i=1}^{N}P_{X}(i) \log P_{X}(i),
\end{equation*}
%and 
%\begin{equation}
%H_Y=-\sum_{j=1}^{N}P_{Y}(j) \log P_{Y}(j),
%\end{equation}
where $P_{X}(i)$ is the probability of a random event $i$ to occur in $X$ (similarly for $H_Y$). The joint entropy, $H_{XY}$, in Eq. \eqref{IXY1} measures the amount of uncertainty in $X$ and $Y$ when taken together and is defined by
\begin{equation}\label{HXY}
H_{XY}=-\sum_{i=1}^{N}\sum_{j=1}^{N}P_{XY}(i,j) \log P_{XY}(i,j),
\end{equation}
where $P_{XY}(i,j)$ is the joint probability of events $i$ and $j$ to occur simultaneously in $X$ and $Y$, respectively.

Equivalently, one can define MI as
\begin{equation}\label{IXY2}
I_{XY}(N)=\sum_{i=1}^{N}\sum_{j=1}^{N}P_{XY}(i,j) \log\left(\frac{P_{XY}(i,j)}{P_{X}(i)P_{Y}(i)}\right).
\end{equation}
This equation provides a measure of the strength of the dependence between $X$ and $Y$, i.e., the amount of information $X$ contains about $Y$ and vice versa \cite{kullback1997}. When MI is zero, i.e., $I_{XY}=0$, the strength of the dependence is null, and thus $X$ and $Y$ are independent variables. This means knowing $X$ does not provide any information about $Y$ and vice versa. MI is a symmetric quantity as $I_{XY}=I_{YX}$ and thus, not suited to study causal effects between $X$ and $Y$.

The computation of $I_{XY}(N)$ requires the calculation of probabilities on a suitably defined 2D probability space generated by $X$ and $Y$ on which a partition based on the $N^2$ events can be defined (see for example Eqs. \eqref{HXY} and \eqref{IXY2}). The probabilities are defined in terms of the frequency of occurrence of events over all events on the 2D probability space and thus, what will be considered as an event is crucial for the definition of the probability space and its partitioning. $I_{XY}$ can be computed for any pair of nodes, $X$ and $Y$, in the same network and can be compared with $I_{XY}$ of other pairs in the same network. However, $I_{XY}$ is not suitable to compare data from different systems as they may exhibit different correlation-decay times and time scales \cite{eckmann1985ergodic, baptista2008finding, pinto2011density}.

There are many methods to compute MI, and all depend on the approach to calculate the probabilities in Eq. \eqref{IXY2}. These are the bin method \cite{moddemeijer1989estimation}, the density-kernel method \cite{moon1995estimation} and the method of the estimation of probabilities from distances between closest neighbours \cite{kraskov2004estimating}. Here, we use the bin method, and particularly, grids of $N^2$ equally sized cells \cite{bianco2016successful}. This method tends to overestimate MI because of the finite length of recorded time-series data, and the finite resolution of non-Markovian partitions \cite{butte2000mutual, steuer2002mutual}. Even though these errors are systematic and always present for any given non-Markovian partition, we will show in Sec. \ref{sec_results} that when it comes to network inference based on MIR, they still allow for correct network inference when combined with statistical significance tests and the false discovery rate method for multiple hypothesis testing.

\subsection{Mutual information rate}

MIR provides a method that bypasses problems associated with the resolution of non-Markovian partitions. In \cite{bianco2016successful}, the authors have shown how to calculate MIR for two, finite-length time-series, independently of the partitions on the probability space. For infinitely long time-series, MIR is theoretically defined as the MI exchanged per unit of time between the random variables $X$ and $Y$ \cite{Shannon1948,Dobrushin1959,Gray1980}
\begin{flalign}\label{MIR1}
\mbox{MIR}_{XY}&=\lim_{N\to\infty}\lim_{L\to\infty}\sum_{i=1}^{L-1}\frac{I_{XY}(i+1,N)-I_{XY}(i,N)}{L}\nonumber \\[1em]
&=\lim_{N}\to\infty\lim_{L\to\infty}\frac{I_{XY}(L,N)-I_{XY}(1,N)}{L}\nonumber \\[1em]
&=\lim_{N\to\infty}\lim_{L\to\infty}\frac{I_{XY}(L,N)}{L},
\end{flalign}
where $I_{XY}(L,N)$ is the MI between $X$ and $Y$ in Eq. \eqref{IXY1}. In this framework, we consider trajectories of length $L$ that follow an itinerary over cells on a grid of infinitely many cells $N$. Clearly, $I_{XY}(1,N)/L$ tends to zero in the limit of infinitely long trajectories, i.e., when $L\rightarrow\infty$.

For finite-length time-series $X$ and $Y$, the definition in Eq. \eqref{MIR1} can be further simplified, as demonstrated in \cite{bianco2016successful}, to
\begin{equation}\label{MIR_def2}
\mbox{MIR}_{XY}(N)=\frac{I_{XY}(N)}{T(N)},
\end{equation}
where $I_{XY}(N)$ is defined as in Eqs. \eqref{IXY1} and \eqref{IXY2}, i.e., it is the MI between $X$ and $Y$, and $N$ is the number of cells in a Markov partition of order $T$. In particular, the probabilities to calculate $I_{XY}$ are calculated in a Markov partition of order $T$ where $T(N)$ is the shortest time for the correlation between $X$ and $Y$ to be lost in the particular Markov partition \cite{bianco2016successful}. In the next section, we discuss $T(N)$ in more detail and elaborate on its estimation, showing how it can be approximated for finite-size times-series.

\subsection{Estimation of the correlation-decay time}\label{subsec_estimation_correlation_decay_time}

To calculate MI and MIR for a pair of variables $X$ and $Y$, we define a 2D probability space $\Omega$ generated by $X$ and $Y$. The space $\Omega$ is partitioned into a grid of $N\times N$ equally sized cells following the bin method \cite{moddemeijer1989estimation,bianco2016successful}. The probability of an event $i$ in $X$ is then given
\begin{equation*}
P_{X}(i)=\frac{\mbox{number of data points in column }i}{\mbox{total number of data points in }\Omega},
\end{equation*}
and of an event $j$ in $Y$
\begin{equation*}
P_{Y}(j)=\frac{\mbox{number of data points in row }j}{\mbox{total number of data points in }\Omega}.
\end{equation*}
Similarly, the joint probability can be defined by the ratio of points in cell $(i,j)$ of the same partition in $\Omega$ and is expressed by
\begin{equation*}
P_{XY}(i,j)=\frac{\mbox{number of data points in cell }(i,j)}{\mbox{total number of data points in }\Omega}.
\end{equation*}
Therefore, MI can be calculated from Eq. \eqref{IXY2} for a grid size $N$, and is thus partition-dependent as it results in different values for different $N$.

To ensure there is always a sufficiently large number of data points in the cells of the partition of $\Omega$, we require the average number of points in all occupied cells to be sufficiently larger than the number of occupied cells,
\begin{equation*}
\left \langle N_{0}(N) \right \rangle \geq N_{oc}.
\end{equation*}
Here, $N_{oc}$ is the number of occupied cells and $\left \langle N_{0}(N) \right \rangle$ the average number of points in all occupied cells in $\Omega$.

To compute MIR using Eq. \eqref{MIR_def2}, we assume that an underlying Markov partition exists and estimate the correlation-decay time $T(N)$, which is the time when $X$ and $Y$ lose memory from their initial states. In systems with sensitive dependence on initial conditions, such as in chaotic systems, the system becomes unpredictable after $T(N)$ time. The correlation-decay time $T(N)$ can be calculated in different ways, e.g., by using the Lyapunov exponents of the dynamics, the largest expansion rates \cite{Baptistaetal2012} or the diameter of an associated itinerary graph \cite{bianco2016successful}. These ways exploit the fact that the dynamics in $X$ and $Y$ is chaotic and thus, points expand to the whole extend of $\Omega$ after about $T(N)$ time. In the following, we will focus on the itinerary graph method \cite{bianco2016successful}, and will discuss the details of its calculation and its relation to $T(N)$. The method is computationally fast and allows for successful network inference, as we will show in Sec. \ref{sec_results}. These characteristics makes it a reasonable choice.

\subsection{The itinerary graph method for the estimation of the correlation-decay time}\label{subsec_itinerary_graph_method}

The correlation-decay time $T(N)$ depends on quantities such as Lyapunov exponents and expansion rates \cite{bianco2016successful}, which are computationally expensive \cite{Baptistaetal2012}. Here, we estimate it by using the diameter of an associated itinerary graph $G$ \cite{bianco2016successful}, i.e., by computing the number of iterations it takes points in cells in $\Omega$ to expand and cover it completely. This is a necessary condition to determine the shortest time for the correlation to decay to zero. Following \cite{bianco2016successful}, we calculate $T(N)$ from the diameter of the network $G$, based on the dynamics of points mapped from one cell in $\Omega$ to another, namely, on a network where its connectivity is given by the transition of points from cell to cell in $\Omega$, hence the term itinerary graph.

In particular, $G$ can be constructed as follows: We assume that each equally sized cell in $\Omega$, occupied by at least one point, represents a node in $G$. Following the transition of points from one cell to another (itinerary), we can form the connections between nodes in $G$. Specifically, a link between $i$ and $j$ nodes in $G$ exists if points in $\Omega$ travel from cell $i$ to cell $j$, where $i,j=1,\dots,N^2$, i.e., they are the labels of the cells that partition $\Omega$. If the link exists, the entry in $G$ is equal to $1$, otherwise it is set equal to $0$. Therefore, $G$ is defined as the corresponding binary, $N^2\times N^2$, matrix of zeros and ones. Consequently, we define $T(N)$ as the diameter of $G$, i.e., as the longest of all shortest paths in $G$. This amounts to the shortest distance between the two most distant nodes in the network. In the context of the expansion of points in $\Omega$, this is the minimum number of iterations it takes for points in cells in $\Omega$ to spread and cover it. Hence, the method transforms the calculation of the correlation-decay time $T(N)$ to the calculation of the diameter of $G$. In our work, we use Johnson's algorithm \cite{Johnson} to estimate the diameter of $G$ and, thus to estimate $T(N)$.

Since we are using fixed-size, non-Markovian cells to compute $T(N)$ and MIR, errors will occur, causing a systematic bias toward larger MIR values. This means that MIR is partition-dependent and that it is overestimated. To account for these errors, and to render MIR partition-independent, the authors in \cite{bianco2016successful} came up with two normalisation procedures, one with respect to the grid-sizes $N$ and another one with respect to all pairs $X$ and $Y$. This is because bigger $N$ (smaller-size cells) are less likely to produce significantly different partitions to Markovian as opposed to smaller $N$ (bigger-size cells). Moreover, using the fact that $\mbox{MIR}_{XY}=\mbox{MIR}_{YX}$ and $\mbox{MIR}_{XX}=0$, one can narrow down the number of $X$, $Y$ pairs of nodes from $M^2$ to $M(M-1)/2$, which reduces considerably the computational time by almost a half.

\subsection{Network inference based on MIR and statistical significance tests}\label{ni_MIR_sst}

Here, we adopt a simpler approach and do not normalise MIR as in \cite{bianco2016successful,Gohetal2018}. Even though the errors due to the use of non-Markovian partitions will be present, they still allow for successful network inference when combined with statistical significance tests and the false discovery rate method for multiple hypothesis testing. This involves the use of the itinerary graph method in Subsec. \ref{subsec_itinerary_graph_method} to estimate the correlation-decay time $T(N)$ and Eq. \eqref{MIR_def2} to compute MIR combined with the false discover rate method.

To infer the structure of a network from data, the authors in \cite{bianco2016successful,Gohetal2018} used a threshold $\tau\in [0,1]$ and considered that the pair $X$, $Y$ is connected if the normalised MIR value is bigger or equal than the threshold. If so, the corresponding entry in the inferred adjacency matrix $A^i$ is 1, otherwise it is 0. In that sense, the choice of an appropriate $\tau$ becomes crucial in recovering successfully the structure of the initial connectivity matrix. If $\tau$ is set too high, actual connections might be missed, and if set too low, spurious connections might appear in the inferred network. Usually, there is no a priory knowledge how to best set $\tau$, which is addressed here by using a statistical approach which does not make use of thresholds.

In \cite{rubido2014exact}, the authors came up with a way to overcome this problem by determining $\tau$, first by sorting all normalised MIR values in ascending order, and then by identifying the first $X$, $Y$ pair for which the normalised MIR increases more than $0.1$. This is an increase which accounts for an abrupt change in the normalised MIR values. Then, they set the threshold $\tau$ as the middle value of the normalised MIR for the identified pair and that for the immediately previous pair of nodes. This is based on the observation that there are two main groups of normalised MIR values, i.e., one for the connected and one for the unconnected groups of nodes in the network. However, this classification might not always be possible as no such gaps always appear or more than one gaps may appear (see for example panels (c) and (d) in Fig. 6 in \cite{Gohetal2018} for financial and stock market data).

Here, assuming only access to recorded data and to overcome such difficulties, we introduce a statistical approach that does not require a threshold: For each pair of nodes $X$, $Y$, we compute MIR$_{XY}$ using Eq. \eqref{MIR_def2} and calculate a symmetric matrix of MIR values complemented by the use of the false discovery rate method for multiple hypothesis testing to accept or reject connections.

\subsubsection{Calculation of surrogate data}

The statistical approach followed here amounts to creating a number of surrogate data $N^{SD}$ for each pair $X$, $Y$ and to computing their MIR values using Eq. \eqref{MIR_def2}. Since we want to infer the network structure in datasets, we make use of random or twin surrogate data \cite{Prichard_et_al_1994}, depending on the properties of the original dataset, which is discussed next. In both cases, the surrogate data have the same length and dimensionality with the recorded time-series.

In particular, when producing random surrogate datasets, their entries are uniformly distributed random numbers in $(0,1)$ and when producing twin surrogate datasets following \cite{Prichard_et_al_1994}, the method calculates blocks of surrogate data with the same second order properties as the original dataset by transforming the original data into the frequency domain, randomising simultaneously the phases across the time-series and converting the data back into the time domain. The first method destroys any possible relationship among the original data and the second destroys only the phase relationship in the original data, if there is any, preserving at the same time their correlation.

To decide whether to use random or twin surrogate data, we estimate: (i) the number of statistically significant correlated $X$, $Y$ pairs by computing the correlation coefficients of the original dataset and their $p$-values for testing the hypothesis there is no relationship between $X$ and $Y$ (null hypothesis). If the $p$-values are smaller than 0.05 and the absolute value of the correlation coefficient is bigger than 0.5, then the correlation coefficient is considered statistically significant (correlation metric). (ii) the amount of additive noise in the original data assuming they are composed of smoothly varying functions added to Gaussian noise with mean zero and variance $\sigma^2$. At any point, $X$ can be represented by a low order polynomial, i.e., a truncated local Taylor series approximation. Applying finite differences, they kill off low-order polynomial terms, while producing a correlated series of points, where its variance is unchanged at $\sigma^2$. Thus, the method is estimating the variance of additive noise in the original data (noise metric). (iii) the global phase-synchronisation in the original dataset by computing the order parameter $\rho\in[0,1]$ following \cite{Gomez-Gardenes2010}. Here, $\rho=0$ means complete desynchronisation and $\rho=1$, complete synchronisation (global phase-synchronisation metric). If the mean of the noise metric over the nodes $X$ is bigger than $10^{-5}$ or the number of statistically significant correlated $X$, $Y$ pairs over all pairs is bigger than the estimated global phase-synchronisation metric, random surrogate datasets will be used, otherwise twin surrogate datasets will be used. This means if the original data are noisy or linearly correlated, random surrogate data will be used whereas if the original data are highly phase-synchronised, twin surrogate data will be used. The idea is that in any case, we use surrogate data with similar properties with the original data, removing the cause that leads to potential connectivity among the nodes, testing whether connectivity comes from the dynamics or due to chance.

The number of surrogate data, $N^{SD}$, is determined by setting a desired significance level $p^{sl}$, such that
\begin{equation*}
N^{SD}=\frac{1}{p^{sl}}.
\end{equation*}
For example, if we are interested in a 95\% confidence interval, then $p^{sl}=0.05$ and we will use $N^{SD}=1/0.05=20$ surrogate datasets. We quantify how successful the network inference is by means of ROC curves, using the $p^{sl}$ values $10^{-1}$, $10^{-2}$ and $10^{-3}$ that correspond to $N^{SD}=10$, 100 and 1000 surrogate datasets, respectively. These curves quantify the true positive rate (TPR) and false positive rate (FPR) as a function of the number of time samples. TPR is the ratio between the number of links that are correctly inferred and the number of links that actually exist in the network. On the other hand, FPR is the ratio between the number of links that are incorrectly inferred and the actual number of non-existing links in the network. The ROC curves provide further information about the type of errors made in the inference method. In particular, TPR and FPR are defined by
\begin{equation*}
\begin{aligned}
\mbox{TPR}&=\frac{\mbox{TP}}{\mbox{TP+FN}},\\
\mbox{FPR}&=\frac{\mbox{FP}}{\mbox{FP+FN}},
\end{aligned}
\end{equation*}
where TP are the true positive, FN the false negative and FP the false positive connections in the inferred connectivity matrix when compared with the initial connectivity matrix. Both TPR and FPR take values in $[0,1]$. When plotting FPR on the horizontal axis and TPR on the vertical axis, perfect inference corresponds to the point with $\mbox{TPR}=1$ and $\mbox{FPR}=0$, i.e., to the point at the upper left corner on the ROC plot with coordinates $(0,1)$. Positive FPR means inferred connections that are not present in the initial connectivity matrix (false positives). Thus, for successful inference, one would typically require TPR to be as close as possible to 1 and FPR as close as possible to 0, or ideally TPR to be equal to 1 and FPR equal to 0, which corresponds to perfect inference.

\subsubsection{Statistical approach to test for connectivity and the inferred adjacency matrix}

To test whether nodes $X$, $Y$ are connected due to their dynamics, we start by defining the same significance level $p^{sl}$ for all pairs of nodes and the null hypothesis that they are unconnected. Then, we count the number of pairs of surrogate data their MIR value (obtained test statistic) is bigger or equal than the MIR value of $X$ and $Y$ (observed test statistic) and denote it by $c_{XY}$. We use this to compute the probability $p_{XY}=c_{XY}/N^{SD}$, where $N^{SD}$ is the number of surrogate data. $p_{XY}$ is the probability of obtaining a test statistic that is as or more extreme than the observed one, assuming the null hypothesis is true. For example, if $p_{XY}=0.03$, it would mean that if the null hypothesis is true, there would be a $3\%$ chance of obtaining the observed test statistic or a more extreme one. Since this is a small probability, we would reject the null hypothesis and would say that nodes $X$ and $Y$ are connected.

Since we are conducting multiple comparisons for the $M(M-1)/2$ pairs of nodes, there is an increased probability of accepting false positives or type I errors, i.e., accepting the rejection of true null hypotheses. Equivalently, this would mean that even though $X$ and $Y$ would be unconnected, we would accept them as connected. To account for such errors, we perform a false discovery rate multiple hypothesis testing \cite{Benjamini_et_al_2001} based on the probabilities $p_{XY}$ and a desired false discovery rate of $p^{sl}/100$ to guarantee that the probability of having one or more false positives is $p^{sl}$ or less. The FDR procedure described in \cite{Benjamini_et_al_2001} is accurate for any test dependency structure (e.g., for Gaussian variables with any covariance matrix). FDR controls the false discovery rate of a family of hypothesis tests and is the expected proportion of rejected hypotheses that are mistakenly rejected (i.e., the null hypothesis is actually true for those tests). FDR is generally a somewhat less conservative and more powerful method for correcting for multiple comparisons than the Bonferroni correction that provides strong control of the family wise error rate, i.e., the probability that one or more null hypotheses are mistakenly rejected.

If the output of the application of the FDR method for a pair $X$, $Y$ is 1, then the test statistic that produced the $p_{XY}$ value is statistically significant and the null hypothesis of the test for the pair $X$, $Y$ must be rejected. This means if the null hypothesis is true (i.e., nodes $X$ and $Y$ are unconnected), there would be a $p_{XY}$ (small) chance of obtaining the observed test statistic or a more extreme. In this case, since the null hypothesis must be rejected, nodes $X$ and $Y$ are connected and the corresponding entry in the inferred adjacency matrix $A^i$ is set equal to 1. If the null hypothesis cannot be rejected, then nodes $X$ and $Y$ remain unconnected and the corresponding entry in the inferred adjacency matrix $A^i$ is set equal to 0. Following this approach for all $M(M-1)/2$ pairs of nodes, we compute the inferred, symmetric, binary adjacency matrix $A^i$, where its entries $A^i_{kl}$ are given by
\begin{equation*}
A^i_{kl}= \left\{
\begin{array}{ll}
0, & \mbox{when no connection between nodes $k$ and $l$ has been inferred from the statistical approach}\\
1, & \mbox{when a connection between nodes $k$ and $l$ has been inferred from the statistical approach}\\
\end{array} 
.\right.
\end{equation*}
Consequently, if a connection between nodes $k$ and $l$ in $A^i$ has been inferred, then a connection between nodes $l$ and $k$ is implied (undirected connections). Finally, $A^i$ is compared with the initial connectivity matrix $A^o$ to compute the ROC curves as a function of the number of time samples in the datasets.

\section{Results}\label{sec_results}

We start by applying the proposed methodology to correlated normal-variates data from stochastic processes in the absence of underlying equations of motion, and then, to data from deterministic, chaotic systems of coupled maps and coupled ordinary differential equations (ODEs), which we call chaotic dynamical units. In all cases, we aim to recover the initial connectivity matrices that were used to couple the processes or chaotic dynamical units.

The coupling is given by undirected networks in the form of binary, symmetric, adjacency matrices (initial connectivity matrices) that are compared with those that result from the application of the method proposed in Sec. \ref{sec_theor_backgd} (inferred matrices). By binary matrices, we mean that any two nodes in the network can be either bidirectionally connected, which corresponds to an entry equal to 1 or unconnected, which corresponds to an entry equal to 0 in the matrix. Since the connections are undirected, the adjacency matrices (initial connectivity and inferred ones) are symmetric.

We couple the deterministic, chaotic dynamical units according to the initial connectivity matrices, record the evolution of their dynamics and produce time-series data. We use the recorded time-series in a pairwise fashion and the method proposed in Sec. \ref{sec_theor_backgd} to infer the initial connectivity matrices and quantify how successful the inference is as a function of the number of time-samples. We compute ROC curves, where the horizontal axis is the rate of false positive connections (FPR) and the vertical the rate of true positive connections (TPR). In that sense, perfect inference corresponds to $\mbox{TPR}=1$ and $\mbox{FPR}=0$, i.e., to the point $(0,1)$ on the ROC plane. When $\mbox{TPR}=1$, it means the proposed methodology was able to infer correctly all connections in the initial connectivity matrix, and when $\mbox{FPR}=0$, it means no incorrectly inferred connections were found when compared to the initial connectivity matrix. Finally, we plot TPR and FPR as a function of time-sample sizes to show convergence to the point with coordinates $(0,1)$ on the ROC plane, which corresponds to perfect network inference.

\subsection{Correlated normal-variates data}\label{subsec_cnvd}

We start by applying the methodology in Sec. \ref{sec_theor_backgd} to correlated normal-variates data motivated by situations where data can only be recorded, as the underlying dynamics and equations of motion are not known. For example, data from global financial markets of different countries can be recorded over time and are found to be correlated \cite{Gohetal2018}, however the underlying equations that govern their evolution are not known \cite{junior2015dependency}.

\begin{figure}[!ht]
\centering{
\includegraphics[width=\textwidth,height=0.7\textheight]{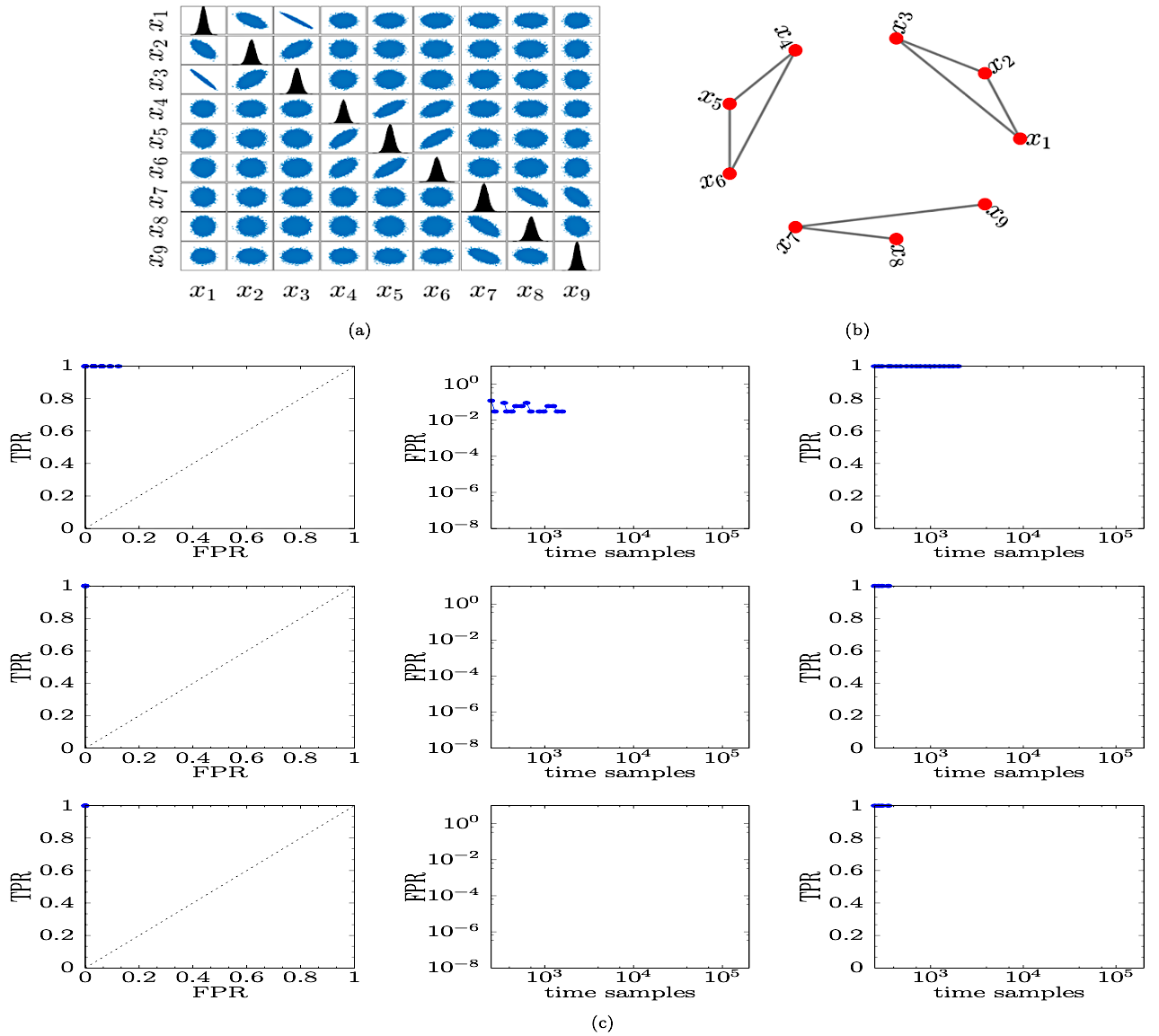}
}
\caption{Application of the proposed methodology to correlated normal-variates data. Panel (A) shows the scatter matrix of the nine datasets that split into the groups $x_1$, $x_2$, $x_3$ (first), $x_4$, $x_5$, $x_6$ (second) and $x_7$, $x_8$, $x_9$ (third). Each group consists of three correlated normal-variates, and the groups are uncorrelated with each other. An ellipsoidal or circular pattern in panel (A) indicates the degree of independence of the datasets, whereas a cigar-shaped cloud of points shows stronger correlation, which can be either positive (correlation) or negative (anti-correlation). The plots in the diagonal of panel (A) show the histograms of the datasets themselves, revealing they are normally distributed. Panel (B) shows the network configuration that results from panel (A), which is the same with the inferred network that resulted from the application of the proposed methodology. Panel (C) shows the ROC curves (left column of plots), FPR vs time samples (middle column) and TPR vs time samples (right column) for $p^{sl}=10^{-1}$, $10^{-2}$ and $10^{-3}$, from top to bottom. Note that the vertical axes in the middle column and, horizontal axes in the middle and right columns of plots in panel (C) are in logarithmic scale and that 0 FPR values in the plots in the middle column of panel (C) are not plotted as the scales in the vertical axes are logarithmic.}\label{fig_cnvd}
\end{figure}

To demonstrate the effectiveness of the method, we used the data of three groups of correlated normal-variates in \cite{Gohetal2018}. Each group $i=1,2,3$ consists of three correlated normal-variates specified by a covariance matrix $\Sigma_i$, where the three groups are uncorrelated with each other. Figure \ref{fig_cnvd}(A) shows the scatter matrix of the three groups. The first group (i.e., $i=1$) consists of variables $x_1$, $x_2$, $x_3$, the second (i.e., $i=2$) of variables $x_4$, $x_5$, $x_6$ and the third (i.e., $i=3$) of variables $x_7$, $x_8$, $x_9$. The covariance matrices $\Sigma_i$ of the three groups are given by
\begin{equation*}
\Sigma_1 =
\begin{bmatrix*}[r]
3.40& -2.75& -2.00 \\
-2.75& 5.50& 1.50 \\
-2.00& 1.50& 1.25
\end{bmatrix*},
\quad
\Sigma_2 =
\begin{bmatrix*}[r]
1.0& 0.5& 0.3\\
0.5& 0.5& 0.3\\
0.3& 0.3& 0.3 
\end{bmatrix*},
\quad
\Sigma_3 =
\begin{bmatrix*}[r]
1.40& -2.75& -2.00\\
-2.75& 5.50& -1.00\\
-2.00& -1.00& 3.25
\end{bmatrix*},
\end{equation*}
respectively, where the subscripts denote the groups.

An ellipsoidal or circular pattern in Fig. \ref{fig_cnvd}(A) indicates the degree of independence of the datasets, i.e., if they are weakly correlated or uncorrelated. A cigar-shaped cloud of points shows stronger correlation, which can be either positive (correlation) or negative (anti-correlation). This is in agreement with the sings of the entries in the covariance matrices. The plots in the diagonal of Fig. \ref{fig_cnvd}(A) show the histograms of the datasets with themselves, from which it results they are normally distributed. For example, one can observe that datasets $x_8$ and $x_9$ are weakly anti-correlated and would not expect to see a connection in the inferred network. In contrast, datasets $x_1$ and $x_3$ are strongly anti-correlated and one would expect to see a connection in the inferred network. Figure \ref{fig_cnvd}(B) shows the initial connectivity network, which is the same with the inferred network that resulted from the application of the proposed methodology. The three groups are apparent in the network in the form of the three disconnected subnetworks.

This is a case where there is a clear distinction between correlated and uncorrelated pairs of nodes as they were constructed per se, what leads to the three disconnected subnetworks in Fig. \ref{fig_cnvd}(B). Here, we used random surrogate data for the computations of the ROC curves in Fig. \ref{fig_cnvd} for $p^{sl}=10^{-1}$, $10^{-2}$ and $10^{-3}$ as the computation of the metrics showed the original data are noisy and correlated, consistent with what is expected from correlated normal-variates data. Figure \ref{fig_cnvd}(C) shows the ROC curves (left column of plots), FPR as a function of time-sample sizes (middle column) and TPR as a function of time-sample sizes (right column) for $p^{sl}=10^{-1}$, $10^{-2}$ and $10^{-3}$, from top to bottom. One can see that for $p^{sl}=10^{-1}$ (top row of plots in panel (C)), FPR fluctuates between $10^{-1}$ and $10^{-2}$ for about $20,000$ time samples and then drops to 0 for bigger time-sample sizes. TPR on the other hand remains fixed at 1 for about $20,000$ time samples showing that all actual connections were inferred correctly. The computations stopped after about $20,000$ time samples as FPR became zero (i.e., no false-positive connections were inferred, as shown in Fig. \ref{fig_cnvd}(B)) and TPR became 1 (i.e., all true positive connections were inferred correctly, as shown in Fig. \ref{fig_cnvd}(B)) for four consecutive time-sample sizes. These results demonstrate that about $20,000$ time samples are enough to successfully infer the network in Fig. \ref{fig_cnvd}(B) for a significance level $p^{sl}=10^{-1}$. A similar conclusion can be drawn for the smaller significance levels $p^{sl}=10^{-2}$ and $10^{-3}$ by looking at the results in the second and third rows of panel (C) (from top to bottom), respectively. In that case, a much smaller size (of about $4,000$) time samples is enough to guarantee perfect network inference, i.e., $\mbox{TPR}=1$ and $\mbox{FPR}=0$. The absence of points in the FPR vs time samples plots (middle column) in panel (C) means that $\mbox{FPR}=0$ as the scale is logarithmic, i.e., no false positive connections were inferred. The results in Fig. \ref{fig_cnvd}(C) show that the proposed methodology inferred successfully the network from correlated normal-variates data. One can see that the data were successfully classified into the three disconnected groups. For example, the connection between variables $x_8$ and $x_9$ is missing as they are weakly anti-correlated and the connection between $x_1$ and $x_3$ is present as they are strongly anti-correlated. This is also corroborated by the results in panel (B) which shows the initial connectivity network with all nodes and connections depicted, being the same with the inferred network that resulted from the application of the proposed methodology. Concluding, we have shown that the proposed methodology can be used to recover the correct number and pairs of correlated normal-variates data and thus, successfully infer the underlying network structure.

\subsection{Network of coupled circle maps}\label{subsec_ccm}

\begin{figure}[!ht]
\centering{
\includegraphics[width=\textwidth,height=0.7\textheight]{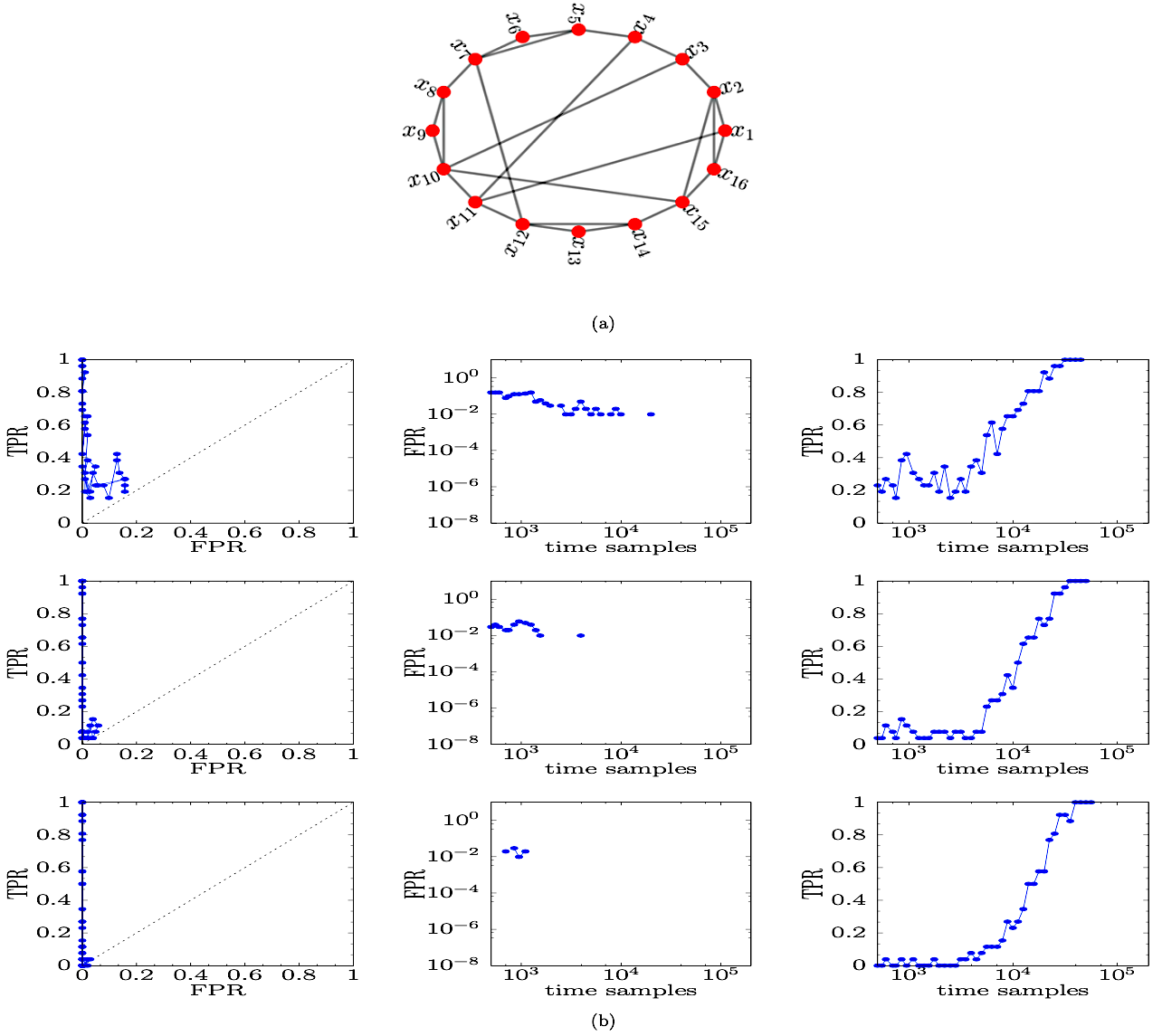}
}
\caption{Application of the proposed methodology to coupled circle maps data for $\alpha=0.03$ and $K=6.9115$. Panel (A) shows the network of the 16 nodes. Panel (B) shows the ROC curves (left column of plots), FPR vs time samples (middle column) and TPR vs time samples (right column) for $p^{sl}=10^{-1}$, $10^{-2}$ and $10^{-3}$, from top to bottom. Note that the vertical axes in the middle column and, horizontal axes in the middle and right columns of plots in panel (B) are in logarithmic scale and that 0 FPR values in the plots in the middle column of panel (B) are not plotted as the scales in the vertical axes are logarithmic.}\label{fig_ccm}
\end{figure}

Next, we applied the method to data obtained from a system of deterministic, chaotic, coupled nonlinear maps and in particular, from coupled circle maps. The network that was used to couple them is shown in Fig. \ref{fig_ccm}(A) and is composed of $M=16$ nodes (circle maps) with dynamics given by \cite{kaneko1990clustering}
\begin{equation}\label{eq_CMN}
x_{n+1}^{i}=(1-\alpha)f(x_{n}^{i},r)+\frac{\alpha }{k_{i}}\sum_{j=1}^{M} A_{ij}f(x_{n}^{j},r),
\end{equation}
where $x_{n}^{i}$ is the $n$th iterate of map $i=1,\ldots,M$ and $\alpha \in[0,1]$ is the coupling strength. $(A_{ij})$ is the binary, symmetric, adjacency matrix of the network shown in Fig. \ref{fig_ccm}(A). The degree of node $i$ is $k_{i}=\displaystyle \sum_{j=1}^{M}A_{ij}$ and $r$ is the parameter of each circle map, which assumes the same value for all $M$ maps. In particular, in Eq. \eqref{eq_CMN}, $f(x_{n},r)$ is the circle map, defined by
\begin{equation}\label{eq_ccme}
f(x_{n},r)=x_{n}+r-\frac{K}{2\pi}\sin(2\pi x_{n}) \mod 1.
\end{equation}

Following \cite{bianco2016successful}, we used $\alpha=0.03$ to build a weakly interacting system of coupled circle maps with $r=0.35$ and $K=6.9115$. These values correspond to fully developed chaos for each circle map (see Eq. \eqref{eq_ccme}). The initial conditions $x_{0}^{i}$ were initialised randomly, drawn form the uniform distribution, in $[0,1]$ and the system was run for a total of $10^6$ iterations. We discarded the first $8\times10^5$ iterations and kept the subsequent $2\times10^5$ iterations to use as the dataset to infer the initial connectivity matrix. Here, we used random surrogate data for the computations of the ROC curves in Fig. \ref{fig_ccm}(B) for $p^{sl}=10^{-1}$, $10^{-2}$ and $10^{-3}$ as the computation of the metrics showed the original data are mainly noisy, what is expected from chaotic datasets.

Figure \ref{fig_ccm}(A) shows the actual network used to generate the data for the coupled circle maps \eqref{eq_ccme} in Eq. \eqref{eq_CMN}. Figure \ref{fig_ccm}(B) shows the ROC curves (left column of plots), FPR as a function of time-sample sizes (middle column) and TPR as a function of time-sample sizes (right column) for the significance levels $p^{sl}=10^{-1}$, $10^{-2}$ and $10^{-3}$, from top to bottom. One can see that for $p^{sl}=10^{-1}$ (top row of plots in panel (B)), FPR fluctuates between $10^{-1}$ and $10^{-2}$ for about $20,000$ time samples after which it drops to 0 for bigger time-sample sizes. TPR on the other hand fluctuates between 0.2 and 0.5 until $50,000$ time samples, after which it increases and reaches 1 at about $30,000$ time samples, showing that all actual connections in the initial connectivity matrix were inferred correctly. The computations stopped after about $50,000$ time samples as FPR became zero (i.e., no false-positive connections were inferred as shown in panel (B)) and TPR became 1 (i.e., all true positive connections were inferred as shown in panel (B)) for four consecutive time-sample sizes. These results demonstrate that about $50,000$ time samples are enough to successfully infer the network shown in panel (B) for $p^{sl}=10^{-1}$. As expected, a similar conclusion can be drawn for the smaller significance levels $p^{sl}=10^{-2}$ and $10^{-3}$ by looking at the results in the second and third rows of panel (B) (from top to bottom), respectively. Again, about $50,000$ time samples are enough for perfect network inference, i.e., $\mbox{TPR}=1$ and $\mbox{FPR}=0$. The absence of points in the FPR vs time samples plots (middle column) in panel (B) for smaller time-sample sizes means that FPR becomes 0 faster than it takes TPR to become 1.

The results in Fig. \ref{fig_ccm}(B) suggest that the proposed methodology inferred successfully the network of deterministic, chaotic, dynamics.

\subsection{Network of coupled logistic maps}\label{subsec_clm}

\begin{figure}[!ht]
\centering{
\includegraphics[width=\textwidth,height=0.7\textheight]{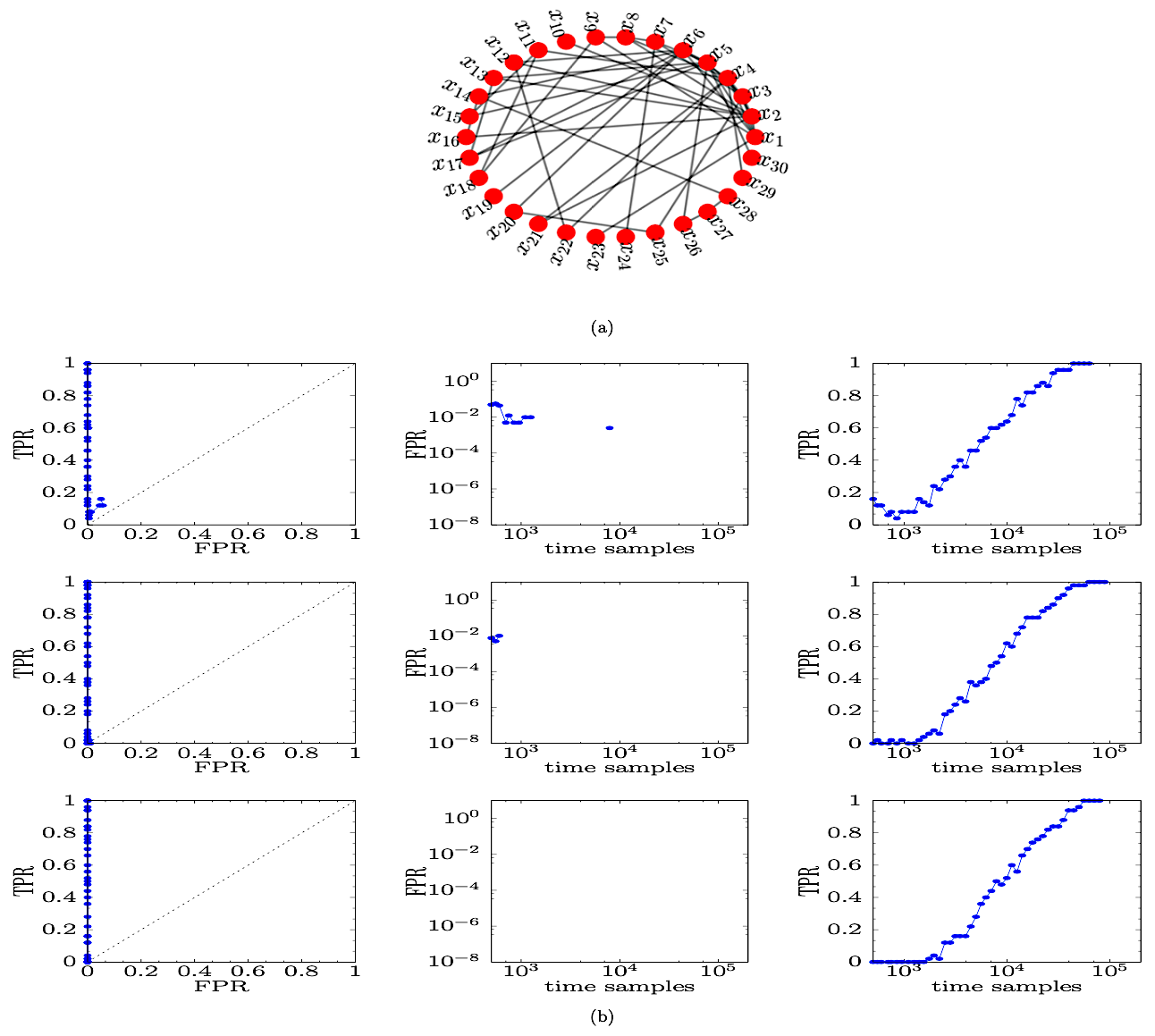}
}
\caption{Application of the proposed methodology to coupled logistic maps data for $\alpha=0.06$. Panel (A) shows the network configuration for the system of 30 coupled logistic maps. Panel (B) shows the ROC curves (left column of plots), FPR vs time samples (middle column) and TPR vs time samples (right column) for $p^{sl}=10^{-1}$, $10^{-2}$ and $10^{-3}$, from top to bottom. Note that the vertical axes in the middle column and, horizontal axes in the middle and right columns of plots in panel (B) are in logarithmic scale and that 0 FPR values in the plots in the middle column of panel (C) are not plotted as the scales in the vertical axes are logarithmic.}\label{fig_clm}
\end{figure}

Here, we apply our methodology to data from coupled logistic maps \cite{bianco2016successful}. Again, they come from a system of deterministic, coupled nonlinear equations that give rise to chaotic dynamics. The network used comprises $M=30$ nodes as shown in Fig. \ref{fig_clm}(A), and we show that the method works for larger-size networks. The dynamics is given by Eq. \eqref{eq_CMN}, whereby each logistic map $f_{i}$ is given by
\begin{equation}\label{eq_clme}
f_{i}(x_{n}^{i},r)=rx_{n}^i(1-x_{n}^{i}),
\end{equation}
where $x_{n}^{i}$ is the $n$th iterate of map $i=1,\ldots,M$ and $r$ the parameter of each logistic map, which is the same for all $M$ maps. In this study, we have used $r=4$ which corresponds to fully developed chaos for each individual logistic map.

Following \cite{bianco2016successful}, we used $\alpha=0.06$ to build a weakly interacting system of coupled, chaotic, logistic maps. The initial conditions $x_{0}^{i}$ were initialised randomly in the interval $[10^{-10},0.3]$ and the system was run for $10^6$ iterations. We discarded the first $8\times10^5$ iterations and kept the next $2\times10^5$ to use as the dataset to infer the network structure, shown in Fig. \ref{fig_clm}(A). We used random surrogate data for the computations of the ROC curves in Fig. \ref{fig_clm}(B), for $p^{sl}=10^{-1}$, $10^{-2}$ and $10^{-3}$ as the computation of the metrics showed the original data are mainly noisy, what is expected from chaotic datasets.

Figure \ref{fig_clm}(A) shows the network of $M=30$ nodes used to produce the data for the coupled logistic maps \eqref{eq_clme} in Eq. \eqref{eq_CMN}. Figure \ref{fig_clm}(B) shows the ROC curves (left column of plots), FPR (middle column) and TPR (right column) as a function of time-sample sizes for $p^{sl}=10^{-1}$, $10^{-2}$ and $10^{-3}$, from top to bottom. One can see that for $p^{sl}=10^{-1}$ (top row of plots in panel (B)), FPR fluctuates between $10^{-1}$ and $10^{-3}$ for the first $1,200$ time samples after which it plummets to 0 as bigger time-sample sizes are considered, with a single exception at $8,000$ time samples where it attains a value between $10^{-2}$ and $10^{-3}$. TPR on the other hand fluctuates between 0 and 0.2 for the first $2,000$ time samples, after which it increases quickly with a linear trend, reaching $\mbox{TPR}=1$ at about $40,000$ time samples, implying that all actual connections in the initial connectivity matrix were inferred correctly. The computations stopped after about $60,000$ time samples as FPR becomes zero (i.e., no false-positive connections were inferred as shown in panel (B)) and TPR becomes 1 (i.e., all true positive connections were inferred as shown in the same panel) for four consecutive time-sample sizes. These results demonstrate that about $40,000$ time samples are enough to successfully infer the network shown in panel (A) for the significance level $p^{sl}=10^{-1}$. As expected, similar conclusions can be drawn for the smaller significance levels $p^{sl}=10^{-2}$ and $10^{-3}$ by looking at the results in the second and third rows of panel (B) (from top to bottom), respectively. Again, about $40,000$ time samples are enough to guarantee perfect network inference of $\mbox{TPR}=1$ and $\mbox{FPR}=0$. The absence of points in the FPR vs time-samples plots (middle column) in panel (B) for smaller time-sample sizes means that $\mbox{FPR}=0$, i.e., no false positive connections were found.

The results in Fig. \ref{fig_clm}(B) suggest that the proposed methodology can be used to successfully infer the network of deterministic dynamics and chaotic datasets, and thus, successfully infer the underlying network structure for larger-size networks.

\subsection{Network of coupled Lorenz systems}\label{subsec_clzs}

Next, we apply the proposed methodology to data obtained from $M=16$ coupled Lorenz systems (see Eqs. \eqref{eq_clzs}), similar to the system studied in \cite{Leguia_et_al_2019}. The authors therein, studied the system in the presence of independent Gaussian noise with zero mean whereas here, we study the same system in the absence of noise as a similar analysis will be presented in Subsec. \ref{subsec_csKpo}. The network is given in Fig. \ref{fig_ccm}(A) and the adjacency matrix $(A_{ij})$ in Eq. \eqref{eq_clzs} is the adjacency matrix of that network.

\begin{figure}[!ht]
\centering{
\includegraphics[width=\textwidth,height=0.7\textheight]{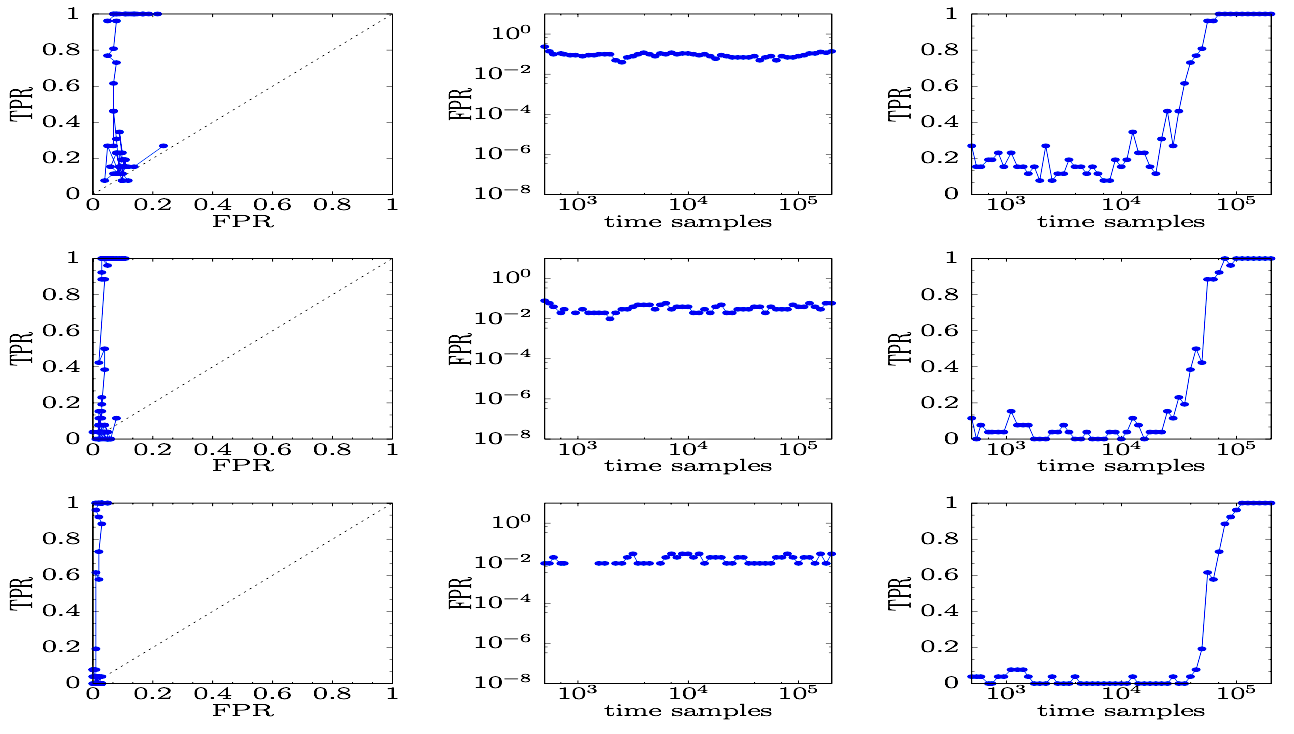}
}
\caption{Application of the proposed methodology to data from $M=16$ coupled Lorenz systems for $K=1.2$. The ROC curves (left column of plots), FPR vs time samples (middle column) and TPR vs time samples (right column) for $p^{sl}=10^{-1}$, $10^{-2}$ and $10^{-3}$, from top to bottom. Note that the vertical axes in the middle column and, horizontal axes in the middle and right columns of plots are in logarithmic scale and that 0 FPR values in the plots in the middle column are not plotted because of the logarithmic scale used in the vertical axes of FPR.}\label{fig_clzs}
\end{figure}

In particular, we considered an undirected network with Lorenz dynamics at the nodes, which are connected via a diffusive coupling with strength $K$ through the $x$ units
\begin{equation*}
\begin{aligned}
\dot{x}_i&=\sigma(y_i-x_i)+K\sum_{j=1}^{M}A_{ij}(x_j-x_i),\\
\dot{y}_i&=x_i(\rho-z_i)-y_i,\label{eq_clzs}\\
\dot{z}_i&=x_iy_i-\beta z_i,
\end{aligned}
\end{equation*}
where $i=1,\ldots,M$, $\sigma=10$, $\rho=28$ and $\beta=8/3$. We have chosen $K=1.2$ as the fixed coupling term for all Lorenz systems as it leads to chaotic dynamics. The initial conditions for $i=1,\ldots,M$ are given by
\begin{equation*}
\begin{aligned}
x_i&= 0.1 + 0.5\xi^x_i,\\
y_i&=-0.19 + 0.5\xi^y_i,\\
z_i&=-0.27 + 0.5\xi^z_i,
\end{aligned}
\end{equation*}
where $\xi^x_i$, $\xi^y_i$ and $\xi^z_i$ are uniformly random numbers in $[0,1]$. The system was solved using the Euler integration (first-order) method with a time step $h=0.01$ and was run up to final integration time $t_f=1,004,000$. Similar results were obtained using the classic Runge-Kutta fourth order integration method. We considered the last $4\times10^5$ data points to use as the dataset to infer the network structure $(A_{ij})$. Here, we used twin surrogate data \cite{Prichard_et_al_1994} for the computations of the ROC curves in Fig. \ref{fig_clzs}, for $p^{sl}=10^{-1}$, $10^{-2}$ and $10^{-3}$ (from top to bottom) as the computation of the metrics showed the original data were more phase correlated than noisy. This method \cite{Prichard_et_al_1994} calculates blocks of surrogate data with the same second order properties as the original time-series dataset by transforming the original data into the frequency domain, randomising the phases simultaneously across the time-series and converting the data back into the time domain. The main effect of the method is that it destroys the phase relationship in the original data (if there is any) preserving at the same time the correlation among the datasets.

Figure \ref{fig_clzs} shows the ROC curves (left column of plots), FPR as a function of time-sample sizes (middle column) and TPR as a function of time-sample sizes (right column) for $p^{sl}=10^{-1}$, $10^{-2}$ and $10^{-3}$, from top to bottom. One can see that for $p^{sl}=10^{-1}$ (top row of plots), FPR fluctuates around $10^{-1}$, which shows the method identified connections as true ones even though they are not present in the initial connectivity matrix, i.e., false positive connections. Also, there is an apparent trend for FPR to increase slightly after about $130,000$ time-samples. TPR on the other hand fluctuates between 0 and 0.4 for the first $30,000$ time samples, after which it increases with a linear trend to $\mbox{TPR}=1$ at about $70,000$ time samples, implying that all actual connections in the initial connectivity matrix were inferred correctly. These results are better appreciated in the TPR vs FPR plot in the first row, where it is clear that the ROC curve tends to stay close to the point $(0,1)$ of perfect inference, even though there are still connections that are falsely identified. The results for $p^{sl}=10^{-1}$ in the first row demonstrate that about $70,000$ time samples are enough to infer all connections in the original network shown in Fig. \ref{fig_ccm}(A), even though at the same time a significant amount of false positive connections is identified. Similar conclusions can be drawn for the smaller significance levels $p^{sl}=10^{-2}$ and $10^{-3}$ by looking at the results in the second and third rows (from top to bottom), respectively. Evidently, about $70,000$ to $100,000$ time samples are enough to infer the original network structure with FPR values decreasing for decreasing significance levels, implying smaller identification errors of false positive connections. The absence of points in the FPR vs time-samples plots (middle column) means that $\mbox{FPR}=0$, i.e., no false positive connections where inferred. The ROC curves approach the vertical axis at $\mbox{FPR}=0$ decreasing by one order of magnitude as the significance level decreases by the same order of magnitude, with TPR approaching 1. Concluding, the quality of the network inference based on MIR improves by one order of magnitude when the significance level decreases by one order of magnitude.

\subsection{Network of coupled stochastic Kuramoto phase oscillators}\label{subsec_csKpo}

In this section, we applied the proposed methodology to data obtained from $M=16$ coupled stochastic Kuramoto phase oscillators (see Eqs. \eqref{eq_csKpo}) \cite{Tirabassi2015}. The network is shown in Fig. \ref{fig_ccm}(A) and the adjacency matrix $(A_{ij})$ in Eqs. \eqref{eq_csKpo} is the adjacency matrix of that network. We consider an undirected network of stochastic Kuramoto phase-oscillators dynamics at the nodes, which are connected through the system of equations
\begin{equation}
d\theta_i=\omega_idt+\frac{K}{M}\sum_{j=1}^{M}A_{ij}\sin(\theta_j-\theta_i)dt+DdW_t^i,\label{eq_csKpo}
\end{equation}
where $\theta_i$ and $\omega_i$ are respectively, the phase and natural frequency of oscillator $i=1,\ldots,M$, and $K$ the fixed coupling constant, which is assumed the same for all oscillators. The term $dW_t^i$ is a Weiner process with 0 mean and variance tuned by the parameter $D$. The distribution of the natural frequencies $\omega_i$ is considered to be centred around 0, since the equations are invariant under the phase translation $\theta_i\rightarrow\theta_j - \langle\omega_i\rangle t$, where $\langle\omega_i\rangle$ is the average natural frequency over all oscillators in the network. Also, it is taken symmetric around 0 \cite{Acebron_et_al_2005} and we assume that the frequencies of the oscillators are uniformly, randomly distributed in $[-2\pi,2\pi]$.

\begin{figure}[!ht]
\centering{
\includegraphics[width=\textwidth,height=0.7\textheight]{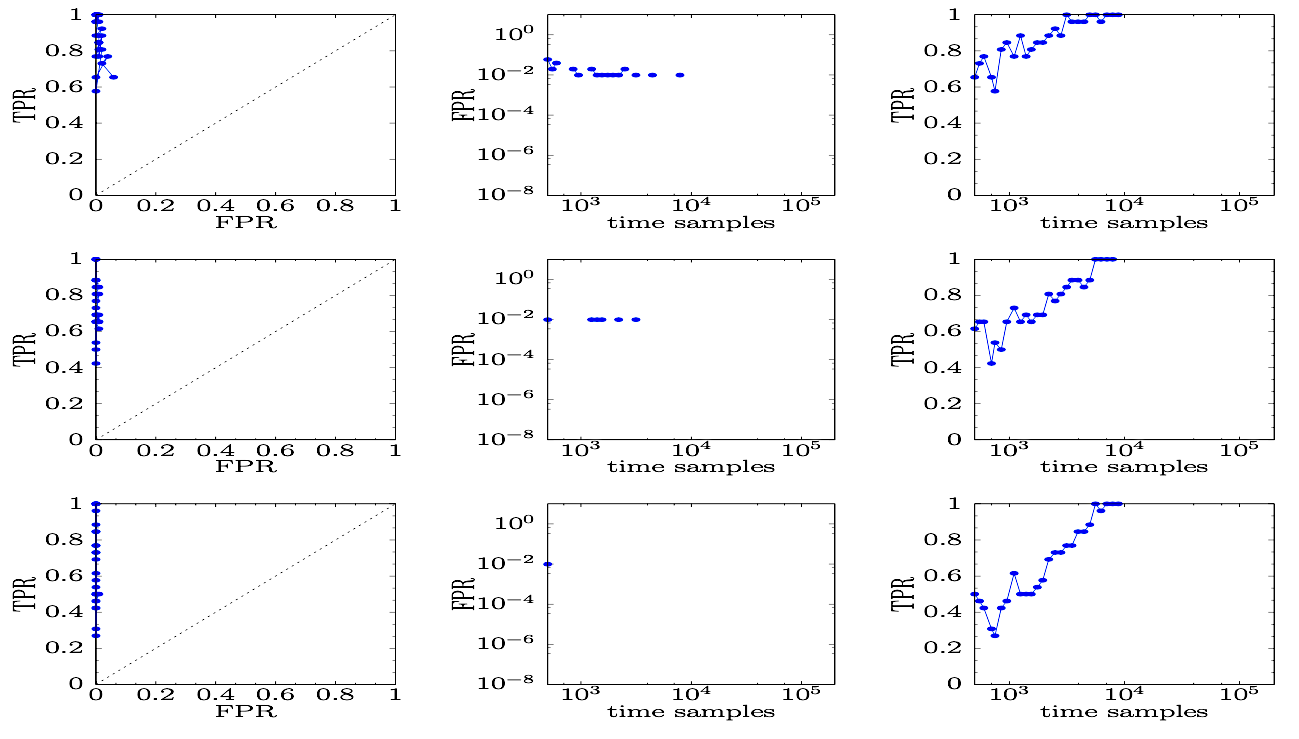}
}
\caption{Application of the proposed methodology to coupled stochastic Kuramoto phase-oscillators data for $K=4$ and $D=0.05$. The ROC curves (left column of plots), FPR vs time samples (middle column) and TPR vs time samples (right column) for $p^{sl}=10^{-1}$, $10^{-2}$ and $10^{-3}$, from top to bottom. Note that the vertical axes in the middle column and, horizontal axes in the middle and right columns of plots are in logarithmic scale and that 0 FPR values in the plots in the middle column are not plotted because of the logarithmic scale used in the vertical axes of FPR.}\label{fig_csKpo}
\end{figure}

The initial conditions used for $i=1,\ldots,M$ are given by
\begin{equation*}
\theta_i=2\pi\xi_i,
\end{equation*}
where $\xi_i$ is a uniformly random number in $[0,1]$. The system was solved by a weak Runge-Kutta method of order 2 with time step $h=0.05$ and was run up to final integration time $t_f=25,000$. We considered the second half $250,000$ data points to use as the dataset to infer the network structure $(A_{ij})$. We used random surrogate data for the computations of the ROC curves in Fig. \ref{fig_csKpo}, for $p^{sl}=10^{-1}$, $10^{-2}$ and $10^{-3}$ (from top to bottom) as the computation of the metrics showed the original data are mainly noisy, what is expected from stochastic dynamics. Following \cite{Tirabassi2015}, for noisy data, one can derive two quantities (or probes) from the phases $\theta_i$ that can be used to infer the structure of the network, namely the instantaneous frequencies $\dot{\theta}_i(t)$ and $Y$-projections $\sin(\theta_i(t))$, i.e., the vertical projections of the unitary vectors of the phases $\theta_i(t)$. Our analysis showed that the quality of inference when using the solution to Eqs. \eqref{eq_csKpo} itself or the $Y$-projections is inferior to that obtained from the instantaneous frequencies. Thus, in Fig. \ref{fig_csKpo}, we present a representative example of network inference for the case $K=4$ and $D=0.05$ using the instantaneous frequencies as a probe.

Figure \ref{fig_csKpo} shows the ROC curves (left column of plots), FPR as a function of time-sample sizes (middle column) and TPR as a function of time-sample sizes (right column) for $p^{sl}=10^{-1}$, $10^{-2}$ and $10^{-3}$, from top to bottom. One can see that for $p^{sl}=10^{-1}$ (top row of plots), FPR fluctuates between $10^{-1}$ and $10^{-2}$ up to $8,000$ time samples, after which it becomes 0. In other words, no more any false positive connections were inferred as the number of time samples increased. TPR on the other hand fluctuates between 0.55 and 0.9 for the first $2,500$ time samples, after which it reaches up to $\mbox{TPR}=1$ at about $3,000$ time samples, implying that all actual connections in the initial connectivity matrix were inferred correctly.

These results are better appreciated in the TPR vs FPR plot in the first row, where it is evident that the points on the ROC curve converge to the point $(0,1)$ of perfect inference. The results for $p^{sl}=10^{-1}$ in the first row demonstrate that about $8,000$ time samples are enough to infer correctly all connections in the original network shown in Fig. \ref{fig_ccm}(A). Similar conclusions can be drawn for the smaller significance levels $p^{sl}=10^{-2}$ and $10^{-3}$ when looking at the results in the second and third rows (from top to bottom), respectively. These findings show that when considering smaller significance levels, the quality of inference improves. The absence of points in the FPR vs time-samples plots (middle column) means that $\mbox{FPR}=0$. The ROC curves approach the vertical axis at $\mbox{FPR}=0$ decreasing by one order of magnitude as the significance level decreases by one order of magnitude, with TPR approaching 1. Concluding, the quality of the network inference improves by one order of magnitude as the significance level decreases by one order of magnitude.

\subsection{Effect of noise on the proposed inference method in networks of coupled stochastic Kuramoto phase oscillators}\label{subsec_enMIRncsKo}

\begin{figure}[!ht]
\centering{
\includegraphics[width=\textwidth,height=0.8\textheight]{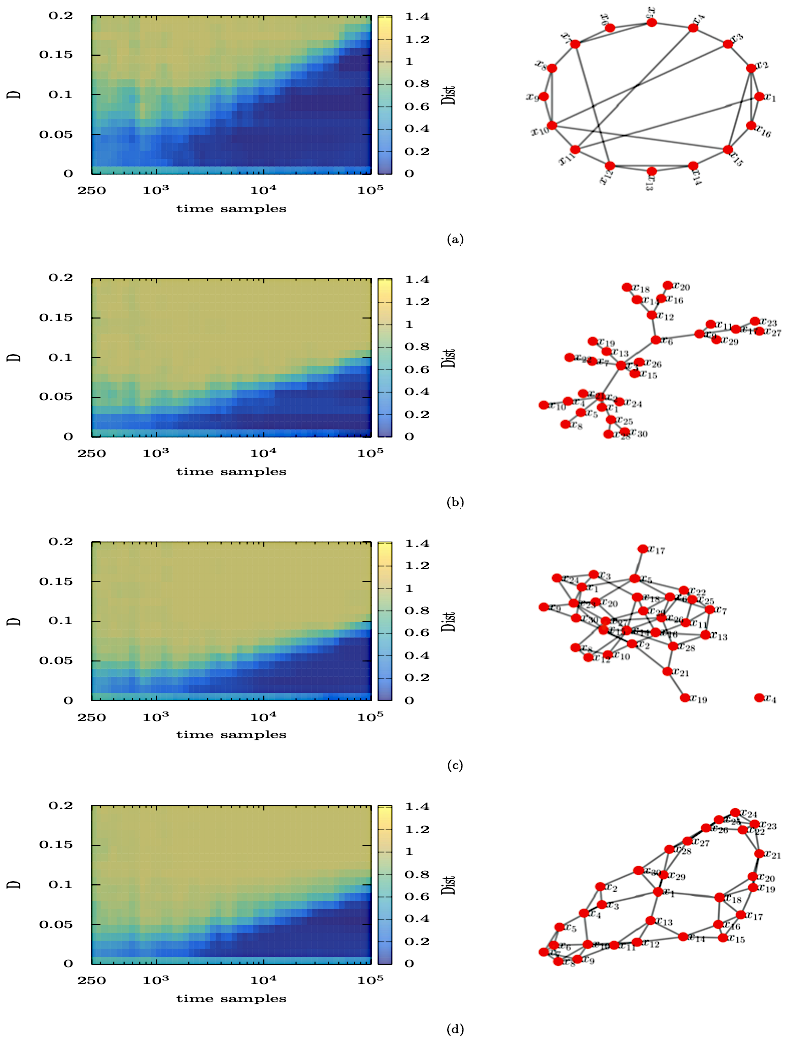}
}
\caption{Effect of noise strength $D$ on the performance of the proposed methodology in: (A) the network of 16 nodes in Fig. \ref{fig_ccm}(A), (B) a scale-free network of $30$ nodes, (C) an Erd\H{o}s-R\'enyi network of 30 nodes with rewiring probability $0.1$, and (D) a small-world network of 30 nodes. Note that in all plots, the horizontal axes are in logarithmic scale and the vertical in linear. The colour bars encode the values of the Euclidean distances between the points on the ROC curves and $(0,1)$ and take values in $[0,\sqrt{2}]$. A zero-distance denotes perfect inference, i.e., the point $(0,1)$ on the ROC plane (not shown).}\label{fig_enMIRncsKo}
\end{figure}

We now move to the study of the effect of additive noise as a function of time-sample sizes on the ability of the proposed methodology to successfully infer the structure of four types of networks. We consider the network of 16 nodes in Fig. \ref{fig_enMIRncsKo}(A), which is the same with the network in Fig. \ref{fig_ccm}(A), the scale-free network in Fig. \ref{fig_enMIRncsKo}(B), the Erd\H{o}s-R\'enyi network with rewiring probability 0.1 in Fig. \ref{fig_enMIRncsKo}(C) and the small-world network in Fig. \ref{fig_enMIRncsKo}(D). In all cases, the time-sample sizes range between $250$ and $100,000$ and the strength of the additive noise $D$ is in $[0,0.2]$. Also, in all cases, the dynamics is given by Eq. \eqref{eq_csKpo} of coupled stochastic Kuramoto phase oscillators with all parameter values and initial conditions as in Subsec. \ref{subsec_csKpo}. We used 10 surrogate datasets to demonstrate the ability of the proposed methodology to successfully infer the structure of the 4 networks for a small such number, which corresponds to $p^{sl}=10^{-1}$. In this study, random surrogate datasets have been used as the metrics showed the original data are mainly noisy, what is expected from stochastic dynamics.

As discussed in Subsec. \ref{subsec_csKpo}, an appropriate probe for coupled stochastic Kuramoto phase oscillators is the instantaneous frequencies, which is used in this analysis. This is also backed by the fact that in experiments or systems in which only a projection of a multidimensional dynamics is available for monitoring, as in brain and climate networks, such projections seem to be a good choice \cite{Tirabassi2015}.

In Fig. \ref{fig_enMIRncsKo}, for each pair of time-sample sizes (horizontal axes) and $D$ values (vertical axes), the proposed methodology for network inference produces a point in the ROC plane by comparing the inferred network with the initial, shown in the left-hand-side plots. We compute the Euclidean distance, denoted by Dist in the figure, between that point and point $(0,1)$ of perfect network inference. These distances are then encoded in the colour bars and lie in $[0,\sqrt{2}]$. The maximum distance $\sqrt{2}$ on the ROC plane corresponds to the lengths of the two diagonals. The smaller the distance, the better the resulting network inference, with perfect inference corresponding to a distance equal to 0, i.e., the adjacency matrices of the inferred and original connectivity matrices are equal.

We plot the Euclidean distance, Dist, for the four networks in Fig. \ref{fig_enMIRncsKo}. Panel (A) shows the results of the analysis for the network of 16 nodes. For small variance of the Wiener process, $D$, as the time-sample size increases, the method infers the structure of the network correctly (blue corresponds to Dist equal to 0). Expectedly, as $D$ increases, larger and larger time-sample sizes are required for successful network inference (as seen in the right-hand-side) of the panel. For $D$ values larger than about 0.2, even larger time-sample sizes would be required for the method to successfully infer the original network, a behaviour reminiscent of the results in \cite{bianco2016successful}.

Apparently, one can arrive at qualitatively similar conclusions when looking at the results in panels (B), (C) and (D) for the scale-free, Erd\H{o}s-R\'enyi and small-world networks, respectively. What is different though, is the fact that for $D$ larger than about 0.1, even larger time-sample sizes would be required for the method to successfully infer the initial connectivity matrix, again reminiscent of the results in \cite{bianco2016successful}.

It is worth it to mention here that scale-free networks, such as the one in panel (B), have heterogeneous topologies in their node-degree distributions, unlike the heterogeneity used in Sec. \ref{subsec_echdMIRnccm} where the connectivities are given by random weights to existing links \cite{rubido2014exact}. In panel (B), for $D=0$, the dynamics is deterministic, and the network is the Barabási scale-free network of 30 nodes in the right-hand-side in the same panel. Hence the 30 Kuramoto models are coupled through this network. As the number of time samples increases, the Euclidean distance, Dist, approaches 0 of successful inference, whereby all connections in the network have been inferred and no incorrectly inferred ones have been found. Particularly, for $10^5$ time samples, the Euclidean distance, Dist, is found to be about 0.077. Hence one would expect the proposed method to work for scale-free networks and, deterministic, chaotic dynamics similarly to the results in Fig. \ref{fig_echdMIRnccm} for random and small-world networks.

Concluding, here we demonstrated the ability of the proposed methodology to infer the structure of the original networks with stochastic dynamics as long as large enough time-series are available for large noise strength $D$, even if only 10 surrogate datasets are used. Based on the results in the previous subsections, had a smaller significance level been used, smaller datasets would be required for perfect network inference, or the performance of the method would be better for the same pairs of noise strengths and time-sample sizes.

\subsection{Effect of coupling heterogeneity degree on the proposed inference method in networks of coupled circle maps}\label{subsec_echdMIRnccm}

Finally, we study the performance of the proposed methodology in networks with varying coupling heterogeneity in their connections which is expected to affect the dynamics of the systems and recorded data \cite{rubido2014exact}. We want to understand how successful the method is in inferring the initial connectivity matrix for Erd\H{o}s-R\'enyi and small-world networks, where their connectivities are given by setting random weights to existing links.

Following \cite{rubido2014exact}, after fixing the initial connectivity matrix by producing a symmetric, binary adjacency matrix $(A_{ij})$ of $M=16$ nodes with rewiring probability 0.3 for both networks, random weights are associated to each connection by considering a new symmetric adjacency matrix with entries
\begin{equation}\label{Wij_matrix}
W_{ij}=A_{ij}(1+g\xi_{ij}),
\end{equation}
for $j>i$, where $0\leq g<1$ is the coupling heterogeneity degree parameter and $\xi_{ij}$ an uncorrelated, zero-mean, uniformly distributed random number in $[-1,1]$. Symmetry in $A$ and $W$ is imposed by considering $A_{ij}=A_{ji}$ and $W_{ij}=W_{ji}$, resulting in undirected connections.

\begin{figure}[!ht]
\centering{
\includegraphics[width=\textwidth,height=0.5\textheight]{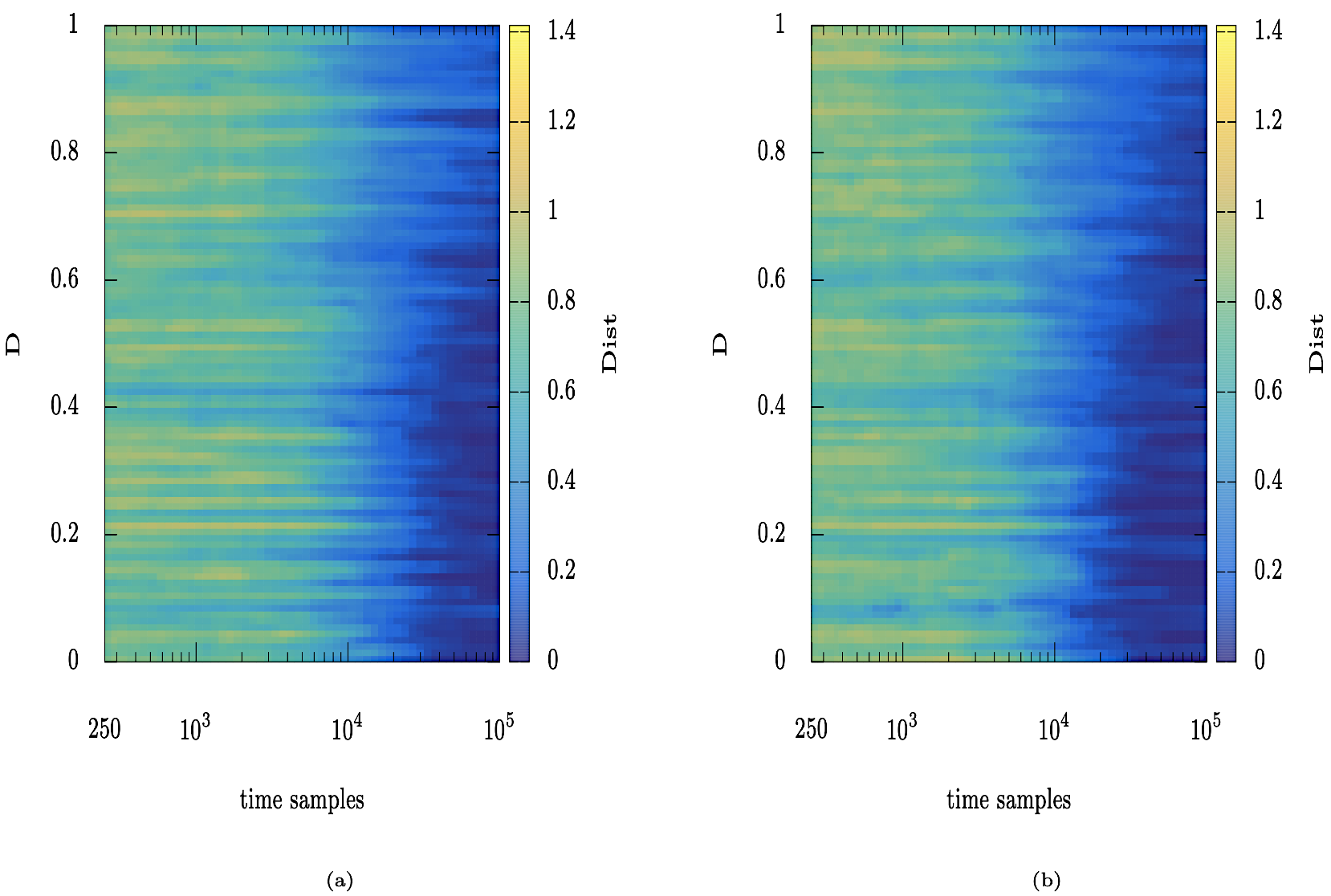}
}
\caption{Effect of coupling heterogeneity degree $g$ on the performance of the proposed methodology in: (A) Erd\H{o}s-R\'enyi networks and (B) small-world networks. In both panels, the networks consist of 16 nodes and the rewiring probability is 0.3. Note that all horizontal axes are in logarithmic scale and all vertical in linear. For each $g$, a different Erd\H{o}s-R\'enyi or small-world network is used. The colour bars encode the values of the Euclidean distances, Dist, between the points on the ROC curves and $(0,1)$, and take values in $[0,\sqrt{2}]$. A zero-distance denotes perfect inference, i.e., the point $(0,1)$ on the ROC plane (not shown).}\label{fig_echdMIRnccm}
\end{figure}

The equation of motion of the $i$-th unit is given by \cite{Kaneko_1992}
\begin{equation}\label{eq_dynamics_ccm_heterogeneity}
x_{n+1}^{i}=(1-\alpha)f(r_i,x_n^i)+\alpha\sum_{j=1}^{M}\frac{W_{ij}}{k_i}f(r_j,x_n^j),
\end{equation}
where, the coupling strength $\alpha=0.1$, $M=16$, $f(r,x)=x+r-1.1\sin(2\pi x)\mod 1$ (i.e., circle maps), $r_i$ is the parameter of the $i$-th circle map, $W_{ij}$ is given by Eq. \eqref{Wij_matrix} and $k_i=\sum_{j=1}^M W_{ij}$ is the weighted degree of node $i$. The parameters $r_i$, $\alpha$ and the underlying topology $(W_{ij})$ determine the dynamics of the circle maps. Here, we set $r_i=r^*-\xi_i\delta r>0$, where $r^*=0.35$ corresponds to the chaotic regime of the circle maps, $\xi_i$ is a random number uniformly distributed in $[0,1]$ and $0\leq \delta r<r^*$ the distribution width that controls the degree heterogeneity in the dynamics. In the following, we considered $\delta r=0.34$ that corresponds to random values $r_i$ in $[0.01,0.35]$.

Next, we obtained the $M=16$ time-series from Eq. \eqref{eq_dynamics_ccm_heterogeneity} for 100, equally spaced values of $g$ in $[0,1]$. For each value, a different network of $M=16$ nodes was used to couple the circle maps, in both cases of network types. Consequently, for each case, 100 different networks of 16 nodes were used, all with rewiring probability 0.3, associated to the 100 equally spaced $g$ values in $[0,1]$. We used 10 surrogate datasets to demonstrate the ability of the proposed methodology to successfully infer the structure of the 2 types of networks for a small such number, which corresponds to $p^{sl}=10^{-1}$. Also, the metrics showed the original data are mainly noisy, what is expected from chaotic dynamics, thus random surrogate datasets were used.

The results of this analysis are shown in Fig. \ref{fig_echdMIRnccm}, where we report the effect of coupling heterogeneity degree $g$ (see vertical axes in linear scale) as a function of the length of the time-series (see horizontal axes in logarithmic scale), on the performance of the proposed methodology for Erd\H{o}s-R\'enyi networks in panel (A) and small-world networks in panel (B). For each pair of time-sample sizes (horizontal axes) and $g$ values (vertical axes), the proposed methodology produced a point on the ROC plane by comparing the inferred network with the initial connectivity matrix $(A_{ij})$. The Euclidean distance, Dist, was computed between that point and point $(0,1)$. The distances are then encoded in the colour bars and lie in $[0,\sqrt{2}]$.

Our findings show that for both types of networks, the proposed methodology inferred successfully the initial connectivity matrix for data produced by dynamics with coupling heterogeneity as a function of time-sample sizes (see, the blue regions on the right hand-side of the plots which correspond to a distance very close to 0). Concluding, the performance of the proposed methodology is robust for coupling heterogeneity degree $g$ as big as 1, for both, Erd\H{o}s-R\'enyi and small-world networks.

\section{Discussion}\label{sec_disc}

Complex systems are the subject of intense research in the last few decades, which has led to the emergence of complexity science. As they consist of many units which may interact in non-trivial ways, and whose aggregate behaviour may be undetermined from the behaviour of the individual units, one can realise their structure by representing them as graphs made up of nodes and edges, i.e., as networks. Thus, recording the evolution of the dynamics on their nodes (i.e., as time-series) is key to inferring their connectivity, in the sense of inferring relations among the time-series, which amounts to finding whether links between pairs of nodes in the networks exist.

This highlights the importance of developing mathematical approaches that can infer network connectivity based on recorded data. Even though, network inference in nonlinear systems, physics, neuroscience, biology, economy, etc., has been studied extensively in recent years using correlation coefficients, cross-correlation, mutual information, Granger causality, rank-based connectivity measures, rank statistics, mutual prediction, phase transfer entropy, etc., it still presents open challenges! Possible reasons might be non-zero values resulting from similarity measures and, nonlinearities and asymmetries in the structure or noise level of the recorded dynamics.

Here, following recent advances in \cite{bianco2016successful,Gohetal2018}, we presented an information-theoretical methodology to infer connectivity in complex systems using time-series data in a pairwise fashion and statistical approaches. The methodology does not depend on the equations of motion, on the processes that produced the data and on the initial connectivity matrix. The only input required is the recorded data. As such, the equations of motions and initial connectivity matrices are only used to record the data and compare the inferred with the initial connectivity matrices by computing receiver operating characteristic curves. The presented method is based on: (a) estimations of the Mutual Information Rate for pairs of time-series and (b) on statistical significance tests for the acceptance of connectivity using the false discovery rate method for multiple hypothesis testing.

The Mutual Information Rate is the rate of information exchanged per unit of time between pairs of time-series and turns out to be an appropriate measure to quantify the exchange of information in systems with correlation. For example, a normalised version of the Mutual Information Rate has been shown to successfully infer the structure of networks in many different cases of toy-dynamical models \cite{bianco2016successful} and, has also been used to infer financial and stock markets data \cite{Gohetal2018}. In \cite{bianco2016successful}, it is shown how to calculate the Mutual Information Rate for Markov partitions, a challenging task as these partitions are difficult or impossible to find. In \cite{rubido2014exact}, two inference methods were compared based on cross-correlation and mutual information. It was shown that when an abrupt change in the ordered values of these measures exists, it is possible to infer correctly the underlying network topology from time-series data. These approaches can be used when well-defined gaps are present. However, they might not be able to infer the correct topology when gaps in the ordered values of the measures are absent or when more than one gaps appear. The presented methodology can be used when only recorded data are available and does not make use of thresholds: an inferred network will be computed based on the recorded data using pairwise computed Mutual Information Rate values and surrogate data to test the null hypotheses for connectivity using the false discovery rate method for multiple hypothesis testing.

The proposed methodology uses the bin or histogram method to compute probabilities for the estimation of the Mutual Information Rate. This is known to lead to overestimation of the Mutual Information Rate for random systems or non-Markovian partitions \cite{steuer2002mutual,Herzeletal1994}, as is the case in this work. The reason is the finite resolution of non-Markovian partitions and the finite length of recorded time-series. These errors are systematic and always present in the computation of the Mutual Information Rate for arbitrary non-Markovian partitions \cite{steuer2002mutual,Herzeletal1994}, e.g., when using the bin method. In \cite{bianco2016successful}, these errors were mitigated by double-normalising the computed Mutual Information Rate values. Here, we showed that this is actually not necessary. Even though, the non-normalised Mutual Information Rate values are overestimated, this is still enough to allow for network inference when combined with the use of significance tests and the false discovery rate method for the acceptance or rejection of connectivity. Comparisons between the MIR values of the dynamics of pairs of nodes can still be made. This is another reason that makes the proposed methodology a reasonable approach to use in network inference from artificial and experimental data.

\section{Conclusions}\label{sec_concl}

In this paper, we presented a methodology that combines information-theoretical and statistical, significance tests using the false discovery rate method for multiple hypothesis testing to infer the structure of networks. We provided the mathematical background on information and network connectivity, and on Mutual Information and Mutual Information Rate. Then, we showed how to estimate the correlation-decay time used in the computations of the Mutual Information Rate and how to combine it with statistical significance tests and the false discovery method to accept those connections that result from the dynamics of the nodes. Following on, we have shown results of network inference for correlated normal-variates data and for coupled circle maps, coupled logistic maps, coupled Lorenz systems and coupled stochastic Kuramoto phase oscillators. Lastly, we have shown the effect of noise on the presented methodology in networks of coupled stochastic Kuramoto phase oscillators and of coupling heterogeneity degree on the presented methodology in networks of coupled circle maps.

We showed that the proposed methodology can infer the correct number and pairs of connected nodes, for correlated normal-variates data and, deterministic and stochastic dynamics. Thus, it can successfully infer the initial connectivity matrix. We also showed that the level of successful network inference improves by an order of magnitude as the significance level decreases by an order of magnitude. This demonstrates the potential of the method in studying scenarios and experimental cases. In the more realistic case of stochastic data, we demonstrated its ability to infer the structure of the initial connectivity matrices as long as large enough time-series are available. Interestingly, had a smaller significance level been used, smaller datasets would be required for perfect network inference, or the performance of the method would be better for the same pairs of noise strengths and time-sample sizes. Moreover, we have shown that the method is able to recover the initial connectivity matrices for Erd\H{o}s-R\'enyi and small-world networks with varying coupling heterogeneity in their connections, which is affecting the dynamics of the systems and thus, the datasets.

Finally, having shown the robust performance of the proposed methodology for different types of computer-generated datasets, dynamics and networks, it would be interesting to study in the future its performance for data coming from experiments or measurements, for example for electroencephalography and magnetoencephalography recordings, financial and stock market data, etc., where the equations of motion are unknown.
 
\section*{Acknowledgments}
We would like to thank Goh Yong Kheng (UTAR) for fruitful discussions and for providing the correlated normal-variates data used in the analysis. The author acknowledges the use of the High Performance Computing Facility (Ceres) and its associated support services at the University of Essex in the completion of this work.

\renewcommand{\baselinestretch}{0.01}
\renewcommand*{\bibfont}{\small}
%\bibliography{Bibliography_5.bib}

\end{document}